\documentclass[twocolumn,showpacs,preprintnumbers,amsmath,amssymb]{revtex4}

\usepackage{graphicx}
\usepackage{dcolumn}
\usepackage{bm}

\begin{document}
\title{Non-local modulation of the energy cascade in broad-band forced turbulence}
\author{Arkadiusz K. Kuczaj} \email{a.k.kuczaj@utwente.nl}
\author{Bernard J. Geurts}
\altaffiliation[]{Also: \emph{Anisotropic Turbulence, Fluid Dynamics
Laboratory, Department of Applied Physics, \mbox{P.O. Box 513},
\mbox{5300 MB Eindhoven}, \mbox{the Netherlands}}}
\affiliation{Multiscale Modeling and Simulation, J.M. Burgers Center
for Fluid Dynamics, NACM, Department of Applied Mathematics,
University of Twente, \mbox{P.O. Box 217}, \mbox{7500 AE Enschede},
\mbox{the Netherlands}}
\author{W. David McComb}
\affiliation{\mbox{School of Physics, University of Edinburgh},
James Clerk Maxwell Building, Mayfield Road, Edinburgh EH9 3JZ,
United Kingdom\\}

\date{\today}

\begin{abstract}

Classically, large-scale forced turbulence is characterized by a
transfer of energy from large to small scales via nonlinear
interactions. We have investigated the changes in this energy
transfer process in {\it broad-band} forced turbulence where an
additional perturbation of flow at smaller scales is introduced. The
modulation of the energy dynamics via the introduction of forcing at
smaller scales occurs not only in the forced region but also in a
broad range of length-scales outside the forced bands due to
non-local triad interactions. Broad-band forcing changes the energy
distribution and energy transfer function in a characteristic manner
leading to a significant modulation of the turbulence. We studied
the changes in this transfer of energy when changing the strength
and location of the small-scale forcing support. The energy content
in the larger scales was observed to decrease, while the energy
transport power for scales in between the large and small scale
forcing regions was enhanced.  This was investigated further in
terms of the detailed transfer function between the triad
contributions and observing the long-time statistics of the flow.
The energy is transferred toward smaller scales not only by
wavenumbers of similar size as in the case of large-scale forced
turbulence, but by a much wider extent of scales that can be
externally controlled.

\end{abstract}


\maketitle

\section{Introduction}

The dynamics of kinetic energy plays a central role in turbulent
flows. The nonlinear term in the Navier--Stokes equations is
responsible for the transfer of energy between any three wavevectors
that form a triad in spectral space~\cite{mccomb:physics:1990}.
Along with the viscous and forcing terms this controls the
production, transfer and dissipation of energy in the system. The
triadic interactions have been studied for decaying and forced
turbulence by many authors (for a review see
\cite{zhou:advances:1998}). Throughout the years various types of
large-scale forcing
methods~\cite{siggia:numerical:1981,kerr:higher:1985,eswaran:examination:1988,chen:high:1992,jimenez:structure:1993,ghosal:dynamic:1995,machiels:predictability:1997,overholt:deterministic:1998,kaneda:energy:2003}
have been proposed to sustain quasi-stationarity in numerical
turbulence as an idealized form of turbulent flow. The aim of such
numerical experiments was to investigate the basic concept of the
Kolmogorov (K41) theory~\cite{kolmogorov:local:1941} that proposes
an inertial range in the kinetic energy spectrum and local transfer
of energy within this range. The turbulent kinetic energy is on
average transferred locally from larger to neighboring smaller
scales.

The purpose of this paper is to numerically investigate the
processes associated with the flow of energy in a turbulent flow.
Specifically, we consider {\it modulated} turbulence in which the
modifications involve the supplementary forcing in a wide range of
modes located in an inertial range of the flow. In the literature, mainly
turbulence with forcing restricted to the large scales has been examined
in detail~\cite{zhou:advances:1998}. The small scale behavior was
found to be energetically quite insensitive to the type of forcing
and at sufficiently high Reynolds numbers a well-developed inertial
range was observed~\cite{zhou:degreees:1993}. Against this
background, we extend the use of forcing methods and investigate
their application directly in the inertial range, thereby focusing
particularly on the competition between transfer and forcing. We
quantify the dominant alterations due to the broad-band forcing in
terms of changes in the energy cascading processes. We pay attention
to the energy transfer function and consider changes that arise in
the contributions from `local', `non-local' and `distant' triadic
interactions. Compared to traditional large-scale forced turbulence,
we observe a strengthening of the contributions of non-local
interactions, leading to a modification of the inertial range
spectrum.

High-resolution direct numerical simulations of turbulence  that
measure the influence of individual terms in the Navier--Stokes
equations on the triadic interactions have been
reported~\cite{domaradzki:analysis:1987,domaradzki:analysis:1988,
domaradzki:local:1990,domaradzki:nonlocal:1992,
domaradzki:energy:1994,ohkitani:triad:1991,waleffe:nature:1991}. It
was found that the energetically dominant triadic interactions
involve sets of three modes in which the magnitude of the wavevector
of one of the modes differs considerably from the other two. This
suggest that statistics of smaller scales may be affected by larger
scales. These dominant processes are not in contradiction with the
Kolmogorov theory because the energy is mainly exchanged between the
two modes of quite similar
wavevector-size~\cite{ohkitani:triad:1991}. Only a small net energy
transfer toward larger wavenumbers arises that involves a detailed
cancellation between many individual triad
transfers~\cite{waleffe:nature:1991}. The spectral space dynamics is
characterized by a multitude of separate transfer-processes among
various modes. These contributions can be collected in pairs with
opposite sign and almost the same magnitude. In total, this leads to
a large number of `near-cancellations' and hence only a comparably
small net effect remains that constitutes the well-known `downward
cascading' toward higher wavenumbers in spectral space. This was
confirmed with the use of helical mode decomposition
in~\cite{waleffe:nature:1991}.

The dynamics of actual turbulent flows seen in nature is usually
characterized by an enormous number of interacting scales, often
perturbed by geometrically complex boundaries and influenced by
additional forces such as rotation and buoyancy. This can lead to
inhomogeneity and anisotropy, which are not covered directly in the
classical view of the Kolmogorov energy cascade and may express
themselves in non-local interactions of various particular scales of
motion. The complexity of such systems motivated us to study in more
detail forcing methods that simultaneously perturb a prescribed
range of scales~\cite{kuczaj:mixing:2006}. Such `broad-band'
agitation of various scales of motion is observed experimentally in
turbulent drag reduction by fibre
suspension~\cite{mccomb:drag:1981,mccomb:laser:1985}, flows through
porous media~\cite{boomsma:metal:2002} and over tree
canopies~\cite{finnigan:turbulence:2000}. In these cases the energy
is transferred abruptly to small scales when the flow reaches an
obstruction. Various other types of flows also exhibit turbulent
motions that coexist at different scales
\cite{pouquet:turbulence:1983}.

\begin{figure}
\begin{center}
\includegraphics[width=0.4\textwidth]{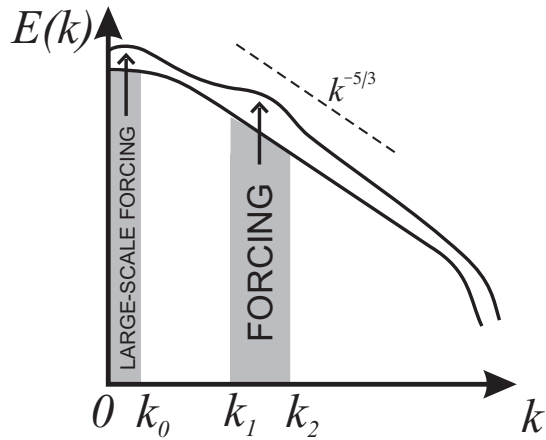}
\end{center}
\caption{Broad-band forcing in spectral space. Large-scale forcing
$k\leq k_0$ with an additional high-$k$ forced region \mbox{$k_1 < k
\leq k_2$}.}\label{fig:bbf}
\end{figure}

To explore the possibilities of a broader application of forcing
methods in turbulence modeling and concurrently examine the energy
dynamics in flows that do not directly follow the classical
Kolmogorov $-5/3$ scaling we employ numerical simulations of
{\it{broad-band}} forced turbulence. The forcing studied in this
paper represents a continual addition/removal of energy from a broad
range of scales in the system, thereby providing the possibility of
altering the characteristic $-5/3$ slope in the kinetic energy
spectrum as predicted by the K41 theory. Specifically, as indicated
in Fig.~\ref{fig:bbf}, we apply the forcing to two regions. The
large-scale forcing $k \leq k_0$ classically agitates the largest
scales in a flow while the additional band $k_1 < k \leq k_2$ is
located in a region of the inertial regime, to allow a direct
competition with the nonlinear transfer term. For inertial-range
scales broad-band forcing introduces explicit energy injection next
to the transfer-term. We varied the spectral support and strength of
the \mbox{high-$k$} band to investigate the modulation of the
turbulence that develops. This distinguishes it from the classical
forcing of large scales only.

In this paper we compute changes in the energy distribution
associated with the broad-band forcing and observe a characteristic
alteration in the spectral energy transfer compared to the classical
Kolmogorov cascading. This alteration expresses itself by additional
local minima and maxima in the transfer function. It is well known
that in cases with large-scale forcing only, negative values are
found for the transfer at the smallest wavenumbers indicating the
energy injection at these scales. The positive values for the
transfer that arise for all other wavenumbers indicates the energy
cascading process to smaller scales. In our case of broad-band
forcing in the inertial range, additional negative regions appear in
the transfer function. These coincide with the additional local
injection of energy. Such a negative region is bordered by nearby
additional maxima in the transfer. These characterize the associated
increased energy transfer to scales just larger or just smaller than
the broad-band forced region.

Forcing applied to different spatial scales simultaneously allows a
non-local modulation of the energy distribution compared to the
reference Kolmogorov case. To quantify the alterations in the energy
transfer we use a decomposition of the velocity field closely
following~\cite{domaradzki:local:1990} and investigate the magnitude
of the contributions from various spatial scales to the overall
energy transfer.

The main finding of this study pertains to the role of broad-band
inertial range forcing in modifying the natural energy cascading
process. This is understood explicitly in terms of changes in the
detailed non-local energy transfer. In addition, we illustrate and
quantify the mechanism of enhancement of the total energy transfer
to smaller scales arising from broad-band forcing and the depletion
of the energy-content in the large scales. Agitation of certain high
wavenumbers can affect well separated low wavenumber components in a
flow. These findings may be relevant for problems that involve the
control of turbulent flow in complex geometries in which various
scales of motions are simultaneously agitated, e.g., in compact
heat-exchangers~\cite{boomsma:metal:2002}. Further applications of
such broad-band forcing may be connected with the observed
modulations of transport properties in physical space leading to an
enhanced scalar dispersion rate~\cite{kuczaj:mixing:2006}.

The organization of this paper is as follows. The mathematical
formulation of the problem is given in Sec.~\ref{sec:model} where
the computational method and the energy transfer terms are also
described. The energy spectra of broad-band forced turbulence and
the modulation of the energy transfer are investigated in
Sec.~\ref{sec:turbul}. In Sec.~\ref{sec:transfer} we present a more
detailed view of the energy transfer processes by computing partitioned
energy transfer function over various spatial scales. The paper
closes with a summary in Sec.~\ref{sec:conc}.

\section{Computational flow model}\label{sec:model}

An overview of the computational model is given in this section
(\ref{eqs}). The forcing method in the broad-band context is
described subsequently (\ref{forc}). In addition, the energy
transfer between different triads is partitioned in a number of
contributions (\ref{trans}) that will be studied numerically in
Sec.~\ref{sec:transfer}. Finally, simulation details are given
(\ref{simu}).

\subsection{Equations of motion} \label{eqs}
The incompressible Navier--Stokes equations in spectral (Fourier)
representation can be written as
\begin{equation}\label{eq:ns}
\left( \partial_t + \nu k^2 \right)u_\alpha ({\bf{k}},t) = \Psi
_\alpha  ({\bf{k}},t) + F_\alpha ({\bf{k}},t),
\end{equation}
where $u_\alpha ({\bf{k}},t)$ is the velocity field coefficient at
wavevector $\bf{k}$ $(k=|\mathbf{k}|)$  and time
$t$~\cite{mccomb:physics:1990}. The non-dimensional kinematic
viscosity $\nu$ is the inverse of the computational Reynolds number
($Re = 1/\nu$). The nonlinear term reads
\begin{equation}\label{eq:psi}
\Psi _\alpha  ({\bf{k}},t) = M_{\alpha \beta \gamma }
\sum\limits_{{\bf{p}} + {\bf{q}} = {\bf{k}}} {u_\beta
({\bf{p}},t)u_\gamma  ({\bf{q}},t)},
\end{equation}
and the forcing term $F_\alpha ({\bf{k}},t)$ is specified in section
\ref{forc}. The tensor $M_{\alpha \beta \gamma}$ in (\ref{eq:psi})
accounts for the pressure and incompressibility effects:
\begin{equation}\label{eq:Mabg}
M_{\alpha \beta \gamma } = \frac{1}{2\imath} \Big(k_\beta D_{\alpha \gamma }
 + k_\gamma  D_{\alpha \beta }\Big),
\end{equation}
in which
\begin{equation}
D_{\alpha \beta } = \delta _{\alpha \beta } - {k_\alpha k_\beta}/{k^2 }.
\end{equation}
Taking the inner product of (\ref{eq:ns}) and $ u_\alpha ^
* ({\bf{k}},t)$, where the asterisk denotes the complex conjugate,
we obtain the energy equation
\begin{equation}\label{eq:e}
\left( \partial_t + 2\nu k^2 \right)E({\bf{k}},t) = T({\bf{k}},t) +
T_{F} ({\bf{k}},t).
\end{equation}
The spectral energy density is denoted by $E({\bf{k}},t) =
\frac{1}{2}u_\alpha ^* ({\bf{k}},t)u_\alpha ({\bf{k}},t)$. The rate
of energy exchanged at wavevector $\bf{k}$ with all other modes in
the system is characterized by the energy transfer function
\begin{equation}\label{eq:t}
T({\bf{k}},t) = u_\alpha ^ * ({\bf{k}},t)\Psi _\alpha ({\bf{k}},t).
\end{equation}
The rate of energy provided by the forcing term is
\begin{equation}\label{eq:tf}
T_{F} ({\bf{k}},t) = u_\alpha ^* ({\bf{k}},t)F_\alpha ({\bf{k}},t),
\end{equation}
and the energy dissipation rate present in (\ref{eq:e}) reads
\begin{equation}\label{eq:eps}
\varepsilon({\bf{k}},t) = 2 \nu k^2 E({\bf{k}},t).
\end{equation}
The three terms $T({\bf{k}},t)$, $T_{F} ({\bf{k}},t)$ and
$\varepsilon({\bf{k}},t)$ represent the energy dynamics in the
system that each typically act in distinct wavenumber regions. The
forcing term $T_{F}(\mathbf{k},t)$ is non-zero in the forced modes
only. In this paper the collection of forced modes will always
contain a low wavenumber band corresponding to large-scale forcing
of the flow. In addition, higher wavenumber contributions will be
included in $T_F(\mathbf{k},t)$. In contrast, the energy dissipation
rate $\varepsilon (\mathbf{k},t)$ is defined in the entire spectral
space, but it is dynamically important primarily for the high
wavenumber range. Finally, the transfer term $T(\mathbf{k},t)$ is
basic to the development of an energy cascade and is a dominant
contribution for wavenumbers in an inertial
range~\cite{mccomb:physics:1990}.

The change of the total energy $\widehat{E}$ in the system is
connected with its viscous dissipation and the total effect of the
forcing. In fact, introducing
\begin{equation}\label{eq:tote}
\widehat{E}(t) = \sum \nolimits_{\mathbf{k}} E(\mathbf{k},t),
\end{equation}
we find
\begin{equation}
\partial_t \widehat{E}(t) =  \widehat{T}_F(t) - \widehat{\varepsilon}(t),
\end{equation}
where $\widehat{\varepsilon}(t) = \sum \nolimits_{\mathbf{k}}
\varepsilon(\mathbf{k},t)$ and $\widehat{T}_F(t) = \sum
\nolimits_{\mathbf{k}} T_F(\mathbf{k},t)$. We used the fact that the
total energy transfer $\widehat{T}(t) =\sum \nolimits_{\mathbf{k}}
T(\mathbf{k},t) =0$. The injection of energy occurs only in the
forced region. This keeps the whole system in a quasi-stationary
state. Normally, the forced region is restricted to the largest
scales in a flow represented by the smallest
wavenumbers~\cite{eswaran:examination:1988,overholt:deterministic:1998}.
The energy introduced in the large scales is transferred to smaller
scales and dissipated primarily in very localized flow-features of
viscous length-scales. By the introduction of an additional source
of energy in the inertial range we will study the perturbation of
the energy cascading process by the forcing. The forcing method
adopted here will be presented next.

\subsection{Forcing method}\label{forc}

Forcing is achieved by applying an additional driving
$F_\alpha(\mathbf{k},t)$ to the velocity field in Fourier space, cf.
(\ref{eq:ns}). Conventionally, the turbulent cascade develops as a
statistical equilibrium is reached, characterized by the balance
between the input of kinetic energy through the forcing and its
removal through viscous dissipation. In literature
(\cite{siggia:numerical:1981,kerr:higher:1985,eswaran:examination:1988,chen:high:1992,jimenez:structure:1993,ghosal:dynamic:1995,machiels:predictability:1997,overholt:deterministic:1998,kaneda:energy:2003}),
we may distinguish several numerical approaches to forced turbulence
that all refer to the agitation of the largest scales of motion.
Here, we modify such classical forcing procedures by allowing for
the simultaneous agitation of a broader range of intermediate-$k$
modes as depicted in Fig.~\ref{fig:bbf}.

We study two ranges of forcing: the classical large-scale forcing
($k \leq k_0$) and small-scale forcing localized in the spectral region
where the transfer of energy $T(\mathbf{k},t)$ is important ($k_1 <
k \leq k_2$). By narrowing or widening the width of the forced
bands, along with a change in their location in spectral space we
can control several aspects of the energy-dynamics. The strength of
forcing is controlled by the amount of energy introduced to various
regions in spectral space.

We expect the small-scale forcing band to influence the inter-scale
energy transfer process not only between scales of similar size but
at a wider spectrum of scales. This may be understood globally as
follows. The process of energy cascading is mainly interpreted via
the resulting local transfer of energy in spectral
space~\cite{ohkitani:triad:1991}. However, this total energy
transfer results from many non-local contributions and these may be
directly altered by the additional small-scale forcing.
Correspondingly an influence on the overall energy cascading process
may occur over an extended wavenumber range. We quantify this effect
by evaluating the nonlinear interactions among the various modes
while they are being perturbed by the broad-band forcing.

In this paper we adopt the recently proposed fractal forcing
\cite{mazzi:fractal:2004}, which involves a power-law dependence of
$F_\alpha$ on the wavenumber:
\begin{equation}\label{eq:mazzi}
F_\alpha  ({\mathbf{k}},t) =  \widetilde{\varepsilon}_{w}
({\mathbf{k}}) \frac{ k^\beta e_\alpha
({\mathbf{k}},t)}{\sum\nolimits_{ {\mathbf{k}} \in {\mathbb{K}} }
{k^\beta} \sqrt {2E({\mathbf{k}},t)}},
\end{equation}
where the coefficient $\beta = D_f - 2$ is connected with the
fractal dimension $D_f$ of the stirrer and
$\widetilde{\varepsilon}_w({\mathbf{k}})$ is the energy input rate
at mode ${\mathbf{k}}$. The set of forced modes $\mathbb{K}$ is
composed of  bands $\mathbb{K}_{m,p}$ ($m \leq p$) which consist of
$p - m + 1$ adjacent spherical shells
$\mathbb{S}_{n}=\frac{2\pi}{L_b}(n-1/2) < |{\mathbf{k}}| \leq
\frac{2\pi}{L_b}(n+1/2)$: $m \leq n \leq p$, in terms of the size of
the computational domain denoted by $L_b$. In the simulations we
always force the first shell $\mathbb{S}_{1}$ and a single high-$k$
band $\mathbb{K}_{m,p}$, if not stated otherwise. The classical
large-scale forcing of the first shell $\mathbb{S}_{1}$ has a
constant energy injection rate $\varepsilon_{w,1}$ in
(\ref{eq:mazzi}) while $\mathbb{K}_{m,p}$ has a constant strength
$\varepsilon_{w,2}$ and a support in spectral space controlled by
$m$ and $p$:
\begin{equation}
\widetilde{\varepsilon}_w ({\mathbf{k}}) = \left\{
\begin{gathered}
  \varepsilon _{w,1} ~~~~{\text{if}}~~~~{\mathbf{k}} \in \mathbb{S}_1,  \hfill \\
  \varepsilon _{w,2} ~~~~{\text{if}}~~~~{\mathbf{k}} \in \mathbb{K}_{m,p},  \hfill \\
  0~~~~~~~~{\text{otherwise}}. \hfill \\
\end{gathered}  \right.
\end{equation}
The vector $\mathbf{e}$ in (\ref{eq:mazzi}) is given by
\cite{mazzi:fractal:2004}:
\begin{equation}
\mathbf{e}(\mathbf{k},t) =
\frac{\mathbf{u}(\mathbf{k},t)}{|\mathbf{u}(\mathbf{k},t)|} + \imath
\frac{\mathbf{k}\times \mathbf{u}(\mathbf{k},t)}{|\mathbf{k}|
|\mathbf{u}(\mathbf{k},t)|}.
\end{equation}
This vector consists of two parts, either parallel or perpendicular
to the vector $\mathbf{u}(\mathbf{k},t)$. In this forcing procedure,
we have control over the energy input rate, the range of forced
modes and the effective geometrical complexity of the stirrer
represented by the fractal dimension.

The summation over all forced modes of $u^{*}_{\alpha}({\bf{k}},t)
F_{\alpha} ({\bf{k}},t)$ yields a total energy input rate given by:
\begin{equation}\label{eq:tran}
\widehat{T}_{F} (t) = \sum \nolimits_{\mathbf{k}} T_{F} ({\bf{k}},t)
=\sum \nolimits_{\mathbf{k}} u^{*}_{\alpha}({\bf{k}},t) F_{\alpha}
({\bf{k}},t)=\varepsilon_{w},
\end{equation}
where  $\varepsilon_{w} = \varepsilon_{w,1} + \varepsilon_{w,2}$.
The energy input leads to a quasi-stationary state described by the
energy equation
\begin{equation}\label{eq:ene} {\partial_t
\widehat{E}(t)} = \varepsilon_w - \widehat{\varepsilon}(t).
\end{equation}
This characterizes the energy dynamics in the system at the most
global level. We observe that this forcing implies a constant energy
injection rate that results in a fluctuating total energy
$\widehat{E}$ and a fluctuating total energy dissipation rate with
mean $\varepsilon_{w}$.

In the next subsection a more detailed description of the energy
dynamics will be given based on the processes that are responsible
for its transfer.

\subsection{Energy transfer}\label{trans}

A detailed investigation of the energy transfer in large-scale
forced turbulence~\cite{ohkitani:triad:1991} shows that the dominant
triadic interactions occur between wavevectors of quite different
lengths. Hence, large-scale forcing may be directly involved in the
dynamics of much smaller scales \cite{yeung:response:1991}. The
interactions are roughly  classified as ``local'' when the sizes of
all wavevectors in a triad are similar, ``non-local'' when the scale
separation is about a factor $10$-$15$ and ``distant'' when the
separation is much larger~\cite{brasseur:interscale:1994}. It was
shown that the transfer of energy reaches maximum values for triads
with two wavevectors of similar size and one with quite different
length~\cite{ohkitani:triad:1991}. Although, the interactions
between triads can be seen mainly as non-local, the dominant net
energy transfer is local, i.e., occurring between similar
scales~\cite{zhou:degreees:1993,
zhou:interacting:1993,zhou:scale:1996}. The interactions produce
forward and backward energy transfer that combined result in a small
net forward energy transfer because of the detailed balance between
contributions that virtually cancel each other
\cite{waleffe:nature:1991}. The forward cascade in the inertial
range was found to be dominated by local and non-local interactions,
while the distant interactions do not significantly transfer
energy~\cite{brasseur:interscale:1994}. All these findings concern
the classical turbulence forced at the largest scales.

Against this background, we ask what the turbulence response will be
to a broad-band perturbation of the energy transfer processes? In
recent literature a somewhat related study was reported in
\cite{suzuki:modification:1999}. Decaying turbulence that starts
from an initial condition with an energetically strongly enhanced
small-scale band of modes was studied. The presence of the extra
small-scale band was found to reduce the intensity of the developing
turbulence by enhancing the {\it{non-local energy cascade}} directly
towards smaller scales. This removes the kinetic energy more
efficiently. The energy feeding mechanisms and energy transfer also
attract much attention in transitional and turbulent flows with an
active control~\cite{moin:feedback:1994}. The modulation induced by
the broad-band forcing has its consequences not only in the spectral
space dynamics of a flow but also in its physical space transport
properties~\cite{kuczaj:mixing:2006}.

\begin{figure}
\begin{center}
\includegraphics[width=0.4\textwidth]{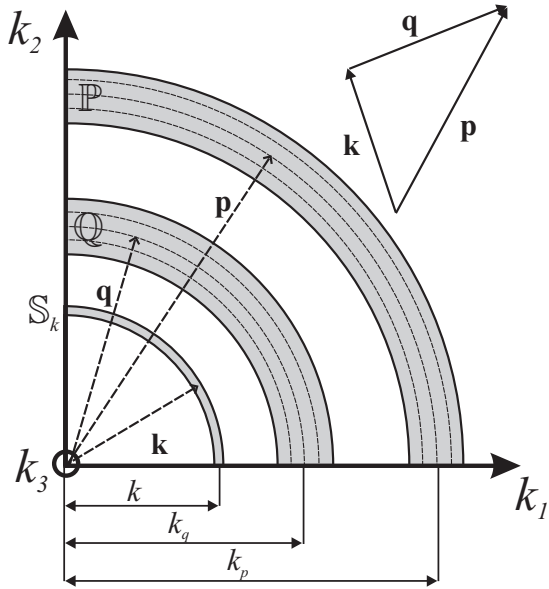}
\end{center}
\caption{Schematic triadic interaction that occurs between
wavevector $\mathbf{k}$ in shell $\mathbb{S}_{k}$ and wavectors
$\mathbf{p}$, $\mathbf{q}$ taken from regions $\mathbb{P}$ and
$\mathbb{Q}$ of spectral space that each consist of four shells with
central wavenumbers $k_p$ and $k_q$.}\label{fig:kpq}
\end{figure}

To analyze the response of turbulence to the additional broad-band
perturbation in more detail we apply previously developed methods
used in the examination of energy transfer in large-scale forced
turbulence~\cite{domaradzki:local:1990}. Referring to
Fig.~\ref{fig:kpq}, the energy transfer between a wavevector
\mbox{${\bf{k}}=(k_1,k_2,k_3)$} and all pairs of wavevectors
$\bf{p}$ and \mbox{$\bf{q}=\bf{k}-\bf{p}$} with $\bf{p}$, $\bf{q}$
chosen in some prescribed regions $\mathbb{P}$ and $\mathbb{Q}$
will be investigated. Such a decomposition allows measuring the
contribution of separate scales to the transfer function
$T(\mathbf{k},t)$. The precise specification requires a few steps
that are presented next. First, we define the truncated velocity
field as
\begin{equation}\label{eq:upq}
u_\alpha ^{(\mathbb{P},\mathbb{Q})} (\mathbf{k},t) = \left\{
\begin{array}{ll}
u_\alpha (\mathbf{k},t) & \rm{if} \, \bf{k} \in \mathbb{P} \, \rm{or} \,  \bf{k} \in \mathbb{Q},\\
0 & \rm{otherwise}.
\end{array}\right.
\end{equation}
Based on this truncated velocity field we may compute the energy
transfer involving the wavevector $\bf{k}$ and all wavevectors
$\mathbf{p}$ and $\mathbf{q}$:
\begin{equation}\label{eq:tpq}
T_{{\mathbb{PQ}}} (\mathbf{k},t) = \left\{\begin{gathered}
    \widetilde{T}_{\mathbb{PP}} (\mathbf{k},t)                        ~~~~~~~~\rm{if}~~~   \mathbb{P} = \mathbb{Q}, \hfill \\
    \frac{1}{2} \Big( \widetilde T_{\mathbb{PQ}} (\mathbf{k},t) - \widetilde{T}_{\mathbb{PP}} (\mathbf{k},t) \hfill \\
         - \widetilde{T}_{\mathbb{QQ}} (\mathbf{k},t) \Big)   ~~~~\rm{if}~~~   \mathbb{P} \ne \mathbb{Q}, \hfill \\
\end{gathered} \right.
\end{equation}
where
\begin{equation}
\widetilde  T_{{\mathbb{PQ}}} (\mathbf{k},t) = u_\alpha ^ *
(\mathbf{k},t)\Psi _\alpha ^{(\mathbb{P},\mathbb{Q})}
(\mathbf{k},t).
\end{equation}
The nonlinear term $\Psi _\alpha ^{(\mathbb{P},\mathbb{Q})}
(\mathbf{k},t)$ is defined by the convolution of the truncated
fields:
\begin{equation}\label{eq:psipq}
\Psi _\alpha ^{(\mathbb{P},\mathbb{Q})} (\mathbf{k},t) = M_{\alpha
\beta \gamma } \sum\limits_{\mathbf{p} + \mathbf{q} = {\bf{k}}}
u_\beta ^{(\mathbb{P},\mathbb{Q})} (\mathbf{p},t) u_\gamma
^{(\mathbb{P},\mathbb{Q})} (\mathbf{q},t),
\end{equation}
where the sum is over all triads with $\mathbf{p} \in \mathbb{P}$
and $\mathbf{q} \in \mathbb{Q}$ such that
$\mathbf{p}+\mathbf{q}=\mathbf{k}$.

For a statistically isotropic, homogeneous turbulence it is
convenient to average over spherical shells in wavevector space. In
addition, in view of the considerable computational effort involved
in computing all interactions between the very large number of
scales present in the flow, we introduced a slight coarse-graining in
terms of the regions $\mathbb{P}$ and $\mathbb{Q}$ as shown in
Fig.~\ref{fig:kpq}. Specifically, it was found adequate to group
together contributions from four adjacent shells. Other more coarse
`groupings' of wavenumbers have been considered in the literature with
the aim of extracting the dominant interaction processes at a
reasonable computational effort. As an example a `logarithmic'
grouping was adopted in~\cite{domaradzki:local:1990} combining
contributions from bands with a width of $2^{k}$. In this paper we
will look at the interactions of four shells $\mathbb{P}$ at
distance $k_p$ (cf. Fig.~\ref{fig:kpq}) with four shells
$\mathbb{Q}$ at distance $k_q$ that contribute to the nonlinear
energy transfer to shell $\mathbb{S}_{k}$ characterized by the
wavenumber~$k$.

In terms of the transfer function $T_{{\mathbb{PQ}}} (\mathbf{k},t)$
we may now define the required spectral transfer functions. The
energy transfer term (\ref{eq:tpq}) gives the exchange of energy by
the triad $(\mathbf{k},\mathbf{p},\mathbf{q})$ where the latter two
wavevectors are specified by the sets $\mathbb{P}$ and $\mathbb{Q}$
and the triangle constraint. Summing over all modes $\mathbf{k}$ in
shell $\mathbb{S}_{k}$ we obtain the exact exchange of energy in the
$k$-th shell between $k$, $k_p$ and~$k_q$:
\begin{equation}\label{eq:Tkpqt}
T_{pq}(k,k_p,k_q,t) = \sum\nolimits_{\mathbf{k}\in \mathbb{S}_{k}}
{T_{\mathbb{PQ}} (\mathbf{k},t)}.
\end{equation}
We refer to $T_{pq}$ as the 'three-mode' transfer. The total energy
transfer function $T(k,t)$ can be computed directly from
(\ref{eq:t}) or as sum of the contributions from (\ref{eq:Tkpqt}):
\begin{equation}\label{eq:Tkt}
  T(k,t) = \sum\nolimits_{k_p} {T_{p}(k,k_p,t)},
\end{equation}
in which the 'two-mode' transfer $T_{p}$ is given by:
\begin{equation}\label{eq:Tkpt}
  T_{p}(k,k_p,t) = \sum\nolimits_{k_q} {T_{pq}(k,k_p,k_q,t)}.
\end{equation}
The individual transfer-terms $T(k,t)$, $T_{p}(k,k_p,t)$ and
$T_{pq}(k,k_p,k_q,t)$ give respectively more detailed
characteristics of the energy transfer. The total transfer $T(k,t)$
expresses the amount of energy transferred from (negative) or to
(positive) shell $\mathbb{S}_{k}$. All three transfer-terms $T$,
$T_{p}$ and $T_{pq}$ will be used to investigate the transfer of
energy in the sequel.

\subsection{Simulation details} \label{simu}

The numerical integration of the Navier--Stokes equations
(\ref{eq:ns}) is done via a four-stage, second-order,
compact-storage, Runge-Kutta method~\cite{geurts:elements:2004}. To
fully remove the aliasing error we applied a method that employs two
shifted grids and spherical truncation \cite{canuto:spectral:1988}.
We consider the canonical problem of forced turbulence in a cubic
box of side $L_{b}$ with periodic boundary conditions. Direct
numerical simulations are characterized by $N^3$ computational
points, where $N$ is the number of grid-points used in each
direction. A detailed description of the simulation setup and the
validation of the numerical procedure can be found
in~\cite{kuczaj:mixing:2006}. The components of the wavevector
$\mathbf{k}$ are $k_\alpha = (2 \pi/L_b) n_\alpha$ where
\mbox{$n_\alpha =0, \pm 1, \pm 2, \ldots, \pm (N/2-1), -N/2$} for
$\alpha=1, 2, 3$. The numerical simulations are defined further by
the size of the domain ($L_{b}$=1), the computational Reynolds
number $Re$ and the energy injection rates to the two distinct bands
($\varepsilon_{w,1}$, $\varepsilon_{w,2}$).

We will study this homogeneous turbulent flow at two different
computational Reynolds numbers, i.e., $Re=1061$ and $Re = 4243$. In
case of homogeneous, decaying turbulence these Reynolds numbers
correspond to $R_{\lambda}=50$ or $100$, in terms of the initial
Taylor-Reynolds number \cite{kuczaj:mixing:2006}. The large-scale
forcing of $\mathbb{S}_{1}$ has an energy injection rate
$\varepsilon_{w,1}=0.15$ that is used as reference case. For all
simulations  the fractal dimension was kept constant and equal to
$D_f=2.6$~\cite{mazzi:fractal:2004}. The smallest length-scale that
should be accurately resolved depends on the size of the box,
viscous dissipation and energy injection rate. Usually it is
required that  $k_{\max} \eta > 1$
\cite{eswaran:examination:1988,yeung:dynamic:1995,mccomb:conditional:2001}
in terms of the Kolmogorov length-scale $\eta$ and the maximal
magnitude of the wavevector $k_{\max}=\pi N/L_b$ that enters the
computations. In our simulations $k_{\max} \eta \gtrapprox 2 $
indicating that the small scales are well resolved.

We consider time-averaged properties of the turbulent flow. For a
function $h$ these are defined by
\begin{equation}\label{eq:ht}
\left\langle {h} \right\rangle_t  = \lim_{t\rightarrow \infty}
\frac{1}{{t - t_0 }}\int\limits_{t_0 }^t {{h(\tau)} d\tau} \approx
\frac{1}{{{\cal{T}} - t_0 }}\int\limits_{t_0 }^{\cal{T}} {{h(\tau)}
d\tau},
\end{equation}
where ${\cal{T}}$ is sufficiently large. We start the averaging at
 $t_0=5$ which corresponds to about $ 10$ eddy-turnover
times for the simulated cases. The final time was taken equal to
${\cal{T}}=30$, so all results are averaged over approximately $50$
eddy-turnover times. The accuracy of this approximation to the
long-time average, measured as the ratio of the standard deviation
and the mean signal is less than $5$\% for all investigated
quantities.

The energy spectra presented in this paper are shell- and
time-averaged. Moreover, we focus on compensated spectra $E_{c}$ in
which we use non-dimensional Kolmogorov units: $E_{c}(k) =
\langle\widehat{\varepsilon}\rangle_t^{-2/3} k^{5/3} \langle
E_{s}(k,t)\rangle_t$ in terms of the shell-averaged spectrum
$E_{s}(k,t) = \sum\nolimits_{\mathbf{k}\in \mathbb{S}_{k}} {
E(\mathbf{k},t)}$. The compensation of the spectrum is not strictly
required to observe the characteristic changes in the energy
distribution, but as it gives more information about the dominant
scales present in a flow it will be used throughout.

\section{Broad-band forced turbulence}  \label{sec:turbul}

To investigate the energy dynamics in broad-band forced turbulence,
first the influence of variation of the strength and location of the
second forced band $\mathbb{K}_{m,p}$ on the global characteristics
and spectrum of the flow is presented (\ref{sec:turbulA}). Then we
examine in more detail the effects of these variations on the energy
distribution and transfer characteristics in the system
(\ref{sec:turbulB}).

\subsection{Energy distribution in forced turbulence} \label{sec:turbulA}

We first concentrate on the  application of the high-$k$ forcing
band  at {\it{different locations}} in spectral space. We apply a
constant energy input rate $\varepsilon_{w,2}=0.15$ to this band.
Simultaneously, the large-scale forcing to the first shell
$\mathbb{S}_{1}$ is $\varepsilon_{w,1}=0.15$. The computational
Reynolds number is $Re = 1061$. We forced the bands
$\mathbb{K}_{p,p+3}$ for $p = 5, 9, 17, 25$. The parameters of these
simulations with some of the statistics are further presented in the
Appendix (Runs~$1$ and $14 - 17$ in Table \ref{tab:pa} and
\ref{tab:da} are concerned here).

\begin{figure}
\begin{center}
\includegraphics[width=0.4\textwidth]{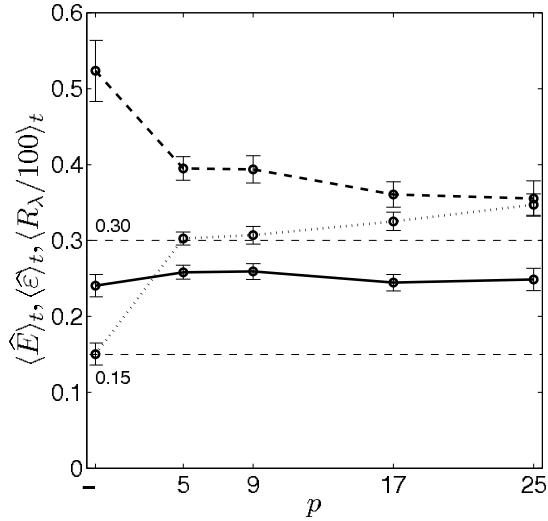}
\end{center}
\caption{Time-averaged total kinetic energy $\widehat E$ (solid),
total energy dissipation rate $\widehat \varepsilon$ (dotted) and
Taylor-Reynolds number $R_\lambda$ (dashed) for forced turbulence
with different locations of the second band at
$Re=1061$.}\label{fig:statb}
\end{figure}

\begin{figure}
\begin{center}
\includegraphics[width=0.4\textwidth]{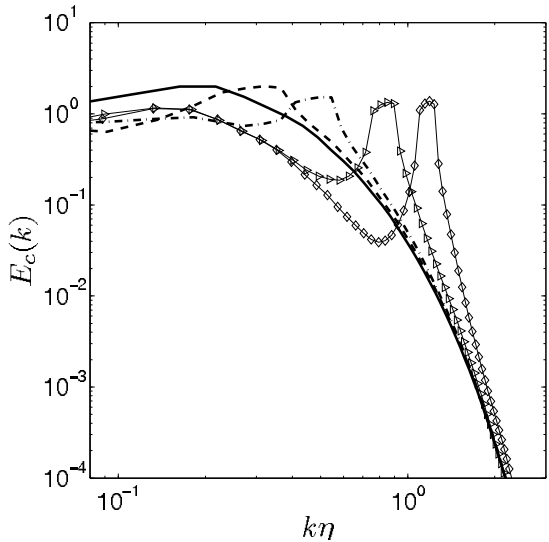}
\end{center}
\caption{Compensated shell- and time-averaged energy spectrum
$E_{c}(k)$ for two-band forcing: $\mathbb{S}_{1}$ and
$\mathbb{K}_{p,p+3}$. Large-scale forcing $\mathbb{S}_{1}$ (solid),
additional forcing in the second band $\mathbb{K}_{5,8}$ (dashed),
$\mathbb{K}_{9,12}$ (dash-dotted), $\mathbb{K}_{17,20}$
($\triangleright$), $\mathbb{K}_{25,28}$ ($\diamond$) at
$Re=1061$.}\label{fig:ebbketa}
\end{figure}

The total kinetic energy, energy dissipation rate and
Taylor-Reynolds number are shown in Fig.~\ref{fig:statb} as a
function of the location of the left-boundary $p$ of the high-$k$
forced band $\mathbb{K}_{p,p+3}$. The first data point refers to the
classical large-scale forcing only (Run~$1$). Application of
broad-band forcing in the different bands changes the
characteristics of the flow modifying primarily the amount of small
scales. This forcing in the second band is seen to increase the
energy dissipation in the system. The Kolmogorov dissipation-scale
and the Taylor-Reynolds number decrease, suggesting that the
characteristic scale at which dissipation plays an important role is
shifted to smaller scales. We notice that the total energy in the
system is only slightly affected by the introduction of forcing.
Moving the broad-band forcing to very small scales implies that
there is no longer a strong influence on the flow because the energy
injected in the small scales appears to be also dissipated
immediately.

The compensated, shell- and time-averaged, energy spectrum for
different locations of the forced region $\mathbb{K}_{p,p+3}$ is
shown in Fig.~\ref{fig:ebbketa}. We may observe that the forcing
causes a non-local depletion in the energy spectrum for the larger
scales while the tail of the spectrum is less affected. The pile up
in the energy spectrum near the forcing region is characteristic of
the explicit high-$k$ forcing and is suggestive of a `blocking' or
reverse cascading. If the separation between $\mathbb{S}_{1}$ and
the high-$k$ band is reduced, then the interaction is stronger and a
considerable depletion of the energy levels in the largest scales
arises. This is in agreement with the large-scale forced turbulence
results, where the local and non-local interactions were found to be
energetically dominant while the distant interactions were mainly
responsible for transferring structural
information~\cite{brasseur:interscale:1994}.

\begin{figure}
\begin{center}
\includegraphics[width=0.4\textwidth]{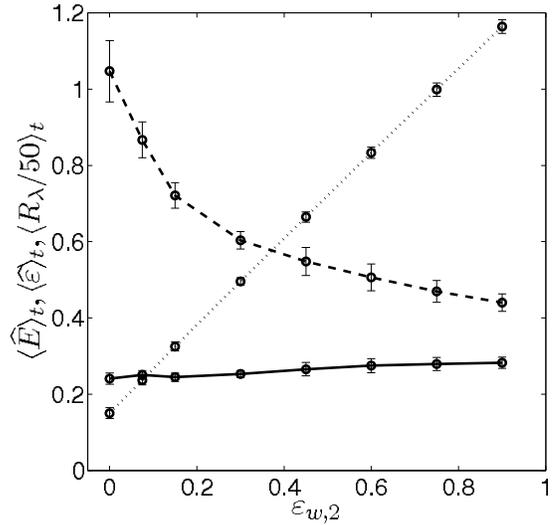}
\end{center}
\caption{Time-averaged total energy $\widehat{E}$ (solid), total
energy dissipation rate $\widehat{\varepsilon}$ (dotted) and
Taylor-Reynolds number $R_\lambda$ (dashed) for two-band forced
turbulence with varying strength in the second band
$\varepsilon_{w,2}$ at $Re=1061$.}\label{fig:stat4}
\end{figure}

\begin{figure}
\begin{center}
\includegraphics[width=0.4\textwidth]{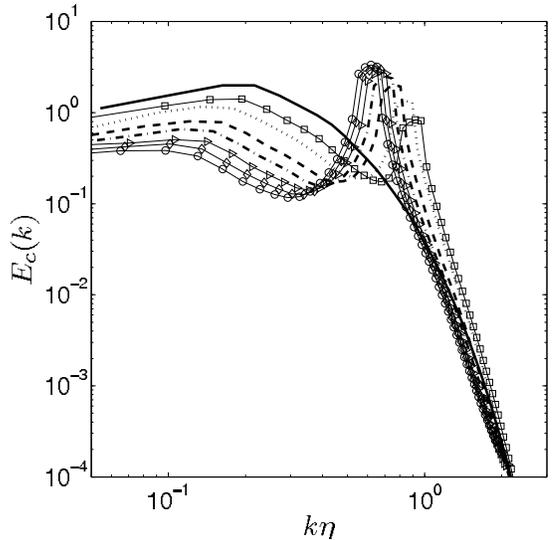}
\end{center}
\caption{Compensated shell- and time-averaged energy spectrum
${E}_{c}(k)$ for forced turbulence in the band $\mathbb{K}_{17,20}$
at different strengths of forcing $\varepsilon_{w,2}$ and $Re=1061$.
Large-scale forcing only (solid), additional second band forcing
with $\varepsilon_{w,2}= 0.07, 0.15, 0.30, 0.45, 0.60, 0.75, 0.90$
denoted as $\square$, dotted, dashed, dash-dotted, $\triangleright$,
$\diamond$, $\circ$, respectively. In each case
$\varepsilon_{w,1}=0.15$.}\label{fig:e4keta}
\end{figure}

An effective modulation of turbulent quantities is possible not only
by a change in the range of forced modes but also via a change in
the energy input rate. To investigate this we adopted an energy
injection rate $\varepsilon_{w,1}=0.15$ for the large-scale forcing
in $\mathbb{S}_{1}$ and we vary the intensity of forcing in the
second band by changing~$\varepsilon_{w,2}$. We adopted the
following values for $\varepsilon_{w,2}$: $0.07$, $0.15$, $0.30$,
$0.45$, $0.60$, $0.75$ or $0.90$ and considered forcing of four or
eight shells in $\mathbb{K}_{17,20}$ or $\mathbb{K}_{17,24}$,
respectively. The parameters and characteristic quantities can be
found in Table \ref{tab:pa} and \ref{tab:da} as Runs $2 - 7$ and $8
- 13$.

\begin{figure}
\begin{center}
\includegraphics[width=0.4\textwidth]{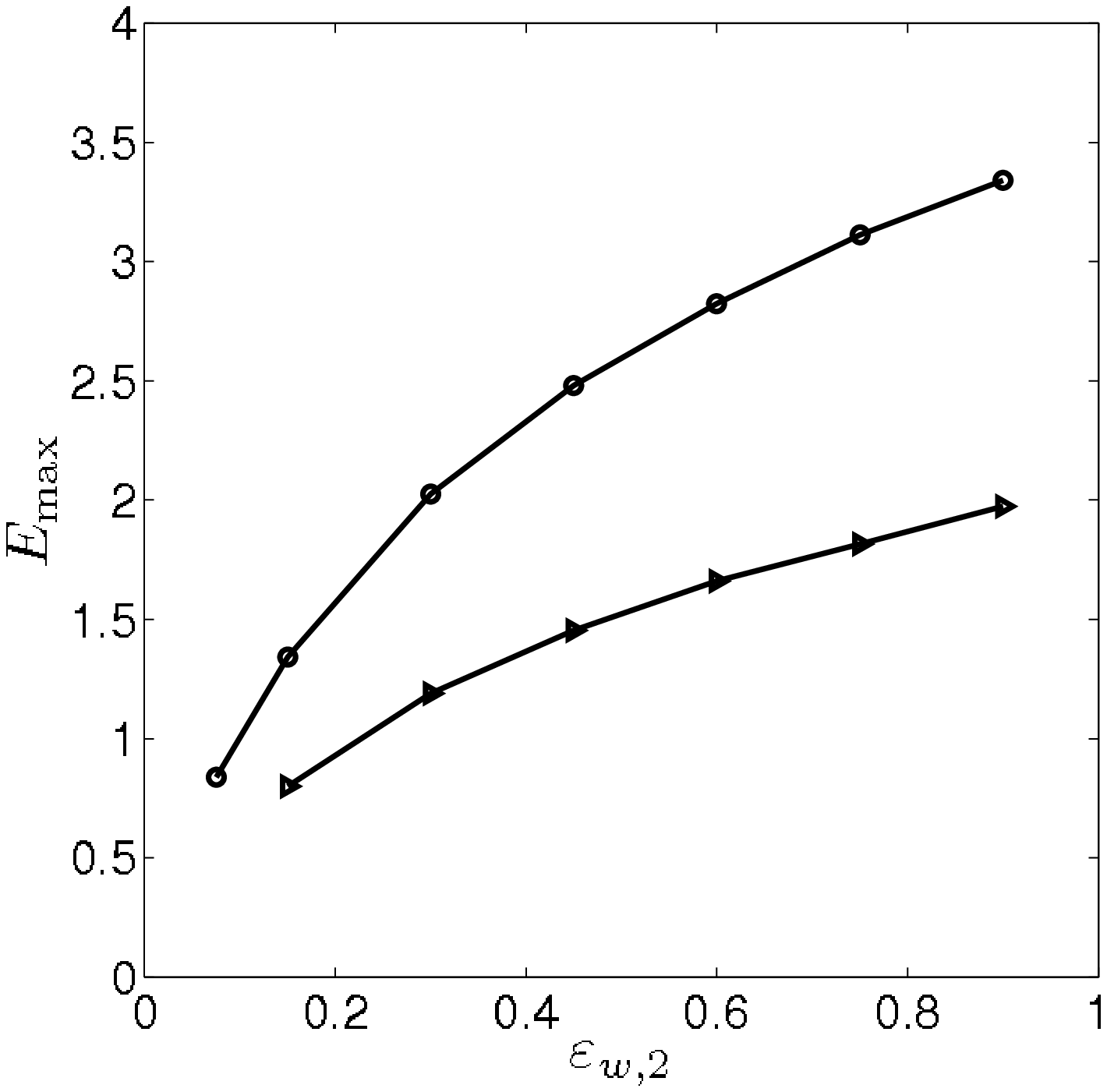}
\end{center}
\caption{Value of the maximum energy $E_{\max }  = \max_{k} \left(
{E_{c}(k)} \right)$ in the forced region for four-shell
$\mathbb{K}_{17,20}$ ($\circ$) and eight-shell $\mathbb{K}_{17,24}$
($\triangleright$) forced turbulence, measured from the compensated
spectra at $Re=1061$.}\label{fig:ketamax1}
\end{figure}

The total energy in the system is only slightly affected by the
forcing-strength in the second band as shown in Fig \ref{fig:stat4}.
An increased forcing strength introduces additional energy into the
flow at small scales that is dissipated very efficiently. This is
expressed by the linear increase in $\langle
\widehat{\varepsilon}\rangle_t$. In Fig.~\ref{fig:e4keta} we present
the compensated energy spectrum for various strengths of the forcing
$\varepsilon_{w,2}$. The energy in the forced region reaches higher
values with increasing $\varepsilon_{w,2}$. This may be further
observed in terms of the energy maximum $E_{\max } = \max_{k} \left(
{E_{c}(k)} \right)$ in Fig.~\ref{fig:ketamax1}. Changing the
strength of the broad-band forcing induces a characteristic
depletion in the larger scales. This suggests that the additional
forcing term enhances the nonlinear interactions, which influence
various scales quite far away from the forced region. The energy
that is injected at the larger scales is transferred even more
effectively through the cascade as $\varepsilon_{w,2}$ increases.
This effect appears similar to the so-called spectral short-cut
observed in nature and experiments~\cite{finnigan:turbulence:2000}.
In case of such a short-cut the energy from larger scales is
diverted quite directly to fine scales largely by-passing the
traditional cascading. This mechanism was explained in the case of flow
over forest-canopies in~\cite{finnigan:turbulence:2000}. We will
investigate it in more detail in the next section.

\begin{figure}
\begin{center}
\includegraphics[width=0.4\textwidth]{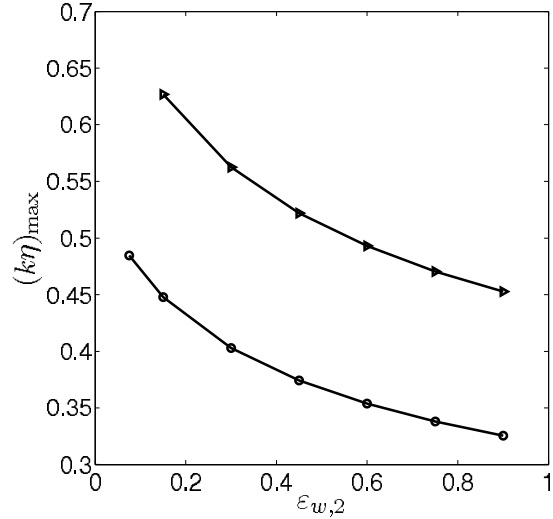}
\end{center}
\caption{Location of the maximum energy in the forced region for
four-shell $\mathbb{K}_{17,20}$ ($\circ$) and eight-shell
$\mathbb{K}_{17,24}$ ($\triangleright$) forced turbulence, measured
from the compensated spectra at $Re=1061$.}\label{fig:ketamax2}
\end{figure}

While the energy-maximum $E_{\max}$ shifts to higher values with
increasing strength of the high-$k$ forcing, the location of the
peak moves towards larger scales. This may be seen from the value of
$(k\eta)_{\max}$ at which the maximum of $E_{c}(k)$ is attained (cf.
Fig.~\ref{fig:ketamax2}). If we inject the same amount of energy to
$\mathbb{K}_{17,24}$ that is two times wider than
$\mathbb{K}_{17,20}$ the maximum of the response decreases (cf.
Fig.~\ref{fig:ketamax1}) while the location of the peak moves
towards smaller scales (cf. Fig.~\ref{fig:ketamax2}). These
numerical experiments are in agreement with observations for
decaying turbulence in which initially additional energy is assigned
to small scales~\cite{suzuki:modification:1999}. The non-local
energy cascade toward the small scales was found to increase
remarkably during the initial period of decay. This is quite similar
to our observed increase of the dissipation rate arising from an
increased forcing of the high-$k$ band.

\begin{figure}
\begin{center}
\includegraphics[width=0.4\textwidth]{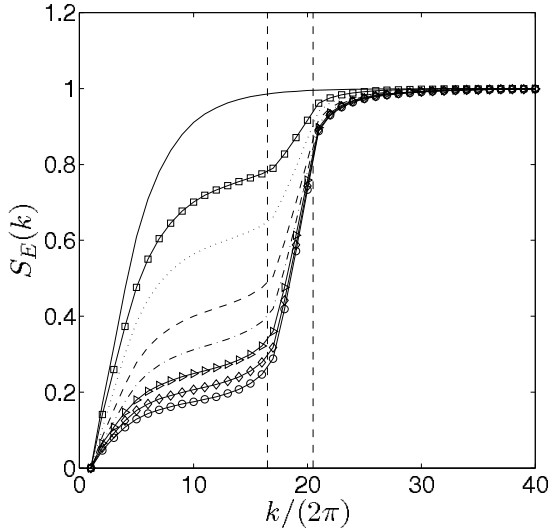}
\end{center}
\caption{Normalized accumulated energy $S_{E}(k)$ for two-band
$\mathbb{S}_{1}$ and $\mathbb{K}_{17,20}$ forced turbulence at
$Re=1061$. Large-scale forcing at $\varepsilon_{w,1}=0.15$ (solid)
with additional second band forcing at $\varepsilon_{w,2}=0.07,
0.15, 0.30, 0.45, 0.60, 0.75, 0.90$ is denoted by the $\square$,
dotted, dashed, dash-dotted, $\triangleright$, $\diamond$ and
$\circ$ curves, respectively.}\label{fig:cume}
\end{figure}

A final quantification of the non-local effect on the spectrum that
arises from the high-$k$ forcing is collected in
Fig.~\ref{fig:cume}. Here we displayed the normalized accumulated
energy
\begin{equation}
S_E (k) = \frac{\sum\nolimits_{k' \leqslant k}
{E_{c}(k')}}{\sum\nolimits_k {E_{c}(k)}}
\end{equation}
in the consecutive shells. As pointed out, varying the properties of
a flow in a specified spectral region can change the behavior of a
flow well outside this region.  In terms of $S_{E}(k)$ we notice
that close to 90\% of the energy is present in the first $10$ shells
(Fig.~\ref{fig:cume}) when only the large-scale forcing is applied.
Influencing the flow at smaller scales in $\mathbb{K}_{17,20}$ is
seen to remove most of the energy from these larger scales while
there is only a slight impact on the dynamics of small scales. This
effect becomes more pronounced with increasing $\varepsilon_{w,2}$.
The underlying changes in the energy transfer will be considered in
more detail in Sec.~\ref{sec:transfer}.

\subsection{Energy transfer spectra}  \label{sec:turbulB}

The transfer of energy in turbulence can be described in spectral
space as interactions of triads of wavevectors ($\mathbf{k}, \mathbf{p},
\mathbf{q}$) that form triangles, i.e.,
$\mathbf{k}=\mathbf{p}+\mathbf{q}$. Direct numerical simulation with
large-scale forcing shows that non-local interactions between
wavevectors combine into a local energy
flow~\cite{zhou:degreees:1993,zhou:interacting:1993}. By applying
forcing that is located  in a high-$k$ range of spectral space we
perturb the `natural' cascading process. The associated changes in
the transfer of energy will be investigated in more detail in this
subsection. Specifically, we focus on  the energy transfer and
energy transport power spectra.

\begin{figure}
\begin{center}
\includegraphics[width=0.4\textwidth]{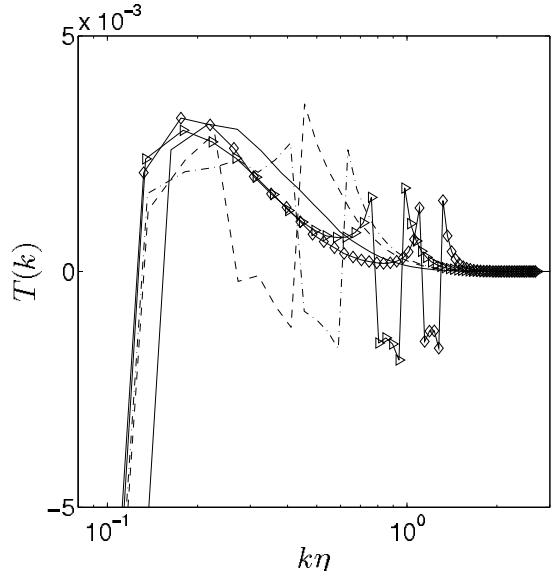}
\end{center}
\caption{Time-averaged energy transfer $T(k)$ for two-band forced
turbulence. Large-scale forcing $\mathbb{S}_{1}$ (solid) with
additional forcing in the $\mathbb{K}_{5,8}$ (dashed),
$\mathbb{K}_{9,12}$ (dash-dotted), $\mathbb{K}_{17,20}$
($\triangleright$), $\mathbb{K}_{25,28}$ ($\diamond$) band at
$Re=1061$.}\label{fig:tbbketa}
\end{figure}

\begin{figure}
\begin{center}
\includegraphics[width=0.4\textwidth]{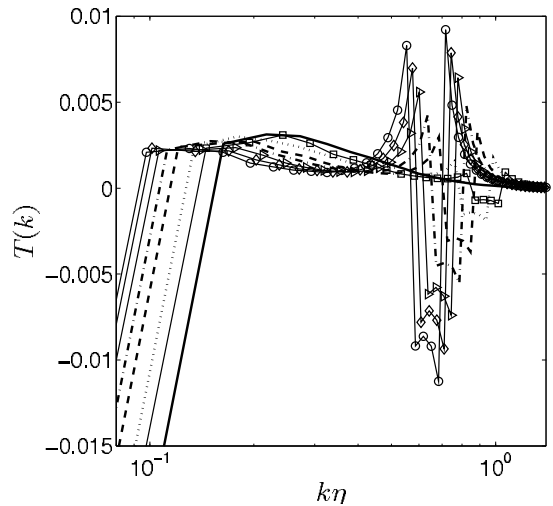}
\end{center}
\caption{Time-averaged energy transfer $T(k)$ for two-band
$\mathbb{S}_{1}$ and $\mathbb{K}_{17,20}$ forced turbulence  for
different strengths of forcing in the second band
$\varepsilon_{w,2}$ at $Re=1061$. Large-scale forcing (solid) with
additional second band forcing at $\varepsilon_{w,2}=0.07, 0.15,
0.30, 0.45, 0.60, 0.75, 0.90$ denoted by $\square$, dotted, dashed,
dash-dotted, $\triangleright$, $\diamond$ and $\circ$ curves,
respectively.}\label{fig:t4k}
\end{figure}

In large-scale forced turbulence energy is injected into the first
shell and removed by the transfer term. This gives rise to negative
values for the energy transfer in the forced region. In the higher
shells the transfer function takes on positive values which
illustrates the transfer of energy through the cascade toward higher
$k$. By invoking the broad-band forcing we influence this basic
energy cascade. This is clearly seen in the energy transfer spectrum
which develops distinctive regions where $T(k) = \langle T(k,t)
\rangle_t$ is negative. In Fig.~\ref{fig:tbbketa} the effect of
variations in the spectral support of the forcing is shown while
Fig.~\ref{fig:t4k} characterizes changes in the transfer function
due to an increased forcing strength of the high-$k$ band. The
transfer function reaches lower values between the low- and high-$k$
forcing regions compared to the large-scale forced case. The reverse
situation appears near the high-$k$ forced band where the transfer
increases with an increase of the forcing intensity. This is in
agreement with the energy spectra presented earlier, where we
observed the depletion of energy between the forced regions. This
effect can be observed more directly from spectra of energy
transport power that will be presented next.

\begin{figure}
\begin{center}
\includegraphics[width=0.4\textwidth]{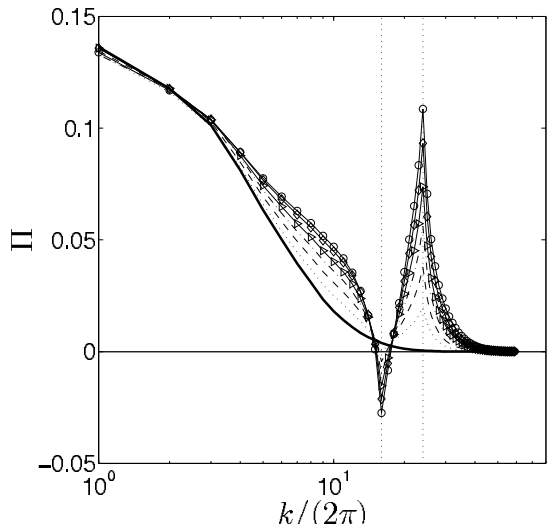}
\end{center}
\caption{Time-averaged transport power spectra for broad-band forced
turbulence in the $\mathbb{K}_{17,24}$ band at $Re=1061$.
Large-scale forcing (solid) with additional second band forced at
$\varepsilon_{w,2}=0.15, 0.30, 0.45, 0.60, 0.75, 0.90$  denoted as
dotted, dashed, dash-dotted, $\triangleright$, $\diamond$, $\circ$
curves, respectively. The supplementary forced region $k_1 < k \leq
k_2$ is denoted with dotted lines.}\label{fig:Pi}
\end{figure}

\begin{figure}
\begin{center}
\includegraphics[width=0.4\textwidth]{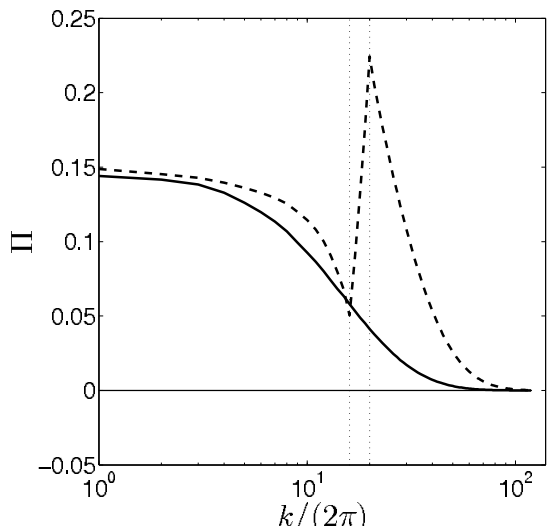}
\end{center}
\caption{Time-averaged transport power spectra for Reynolds number
$Re=4243$ and forcing in the $\mathbb{K}_{17,20}$ band. Large-scale
forcing (solid), additional broad-band forcing in the
$\mathbb{K}_{17,20}$ band with $\varepsilon_{w,2} = 0.30$
(dashed).}\label{fig:Pi1}
\end{figure}

The energy transport power gives the rate at which energy is
transferred from shells $ k'< k$ to those with~\mbox{$k' > k$}:
\begin{equation}\label{eq:pik}
\Pi (k,t)   = \int\limits_{k }^{k_{\max}}{T(k',t)dk'} = -
\int\limits_0^{k} {T(k',t)dk'},
\end{equation}
where \mbox{$k_{\max} = \pi N/L_b$} is the cut-off wavenumber.
We~present the time-averaged transport power spectrum $\Pi(k) =
\langle \Pi(k,t) \rangle_t$ in Fig.~\ref{fig:Pi} for forcing with
various strengths in the $\mathbb{K}_{17,24}$ band. In case of
large-scale forcing only, the transport power is positive for all
$k$ as the energy is transferred toward smaller scales and reaches
zero for large $k$ indicating the general property of the total
transfer function $\widehat{T}(t)=0$. The application of high-$k$
forcing for $k_{1} < k \leq k_{2}$ changes this well-known picture.
First, we note that the values of the transport power are all
similar in the largest scales, where the flow is governed by the
same energy input. The transport power for $0 < k \leq k_{1}$
becomes larger at higher $\varepsilon_{w,2}$. A striking change of
the behavior arises for $k$ near and inside the high-$k$ forced
region. The transport power spectrum even assumes negative values for $k
\approx k_{1}$.

The observed behavior of the transport power in Fig.~\ref{fig:Pi} is
partly due to the relatively low Reynolds number that was used. At
sufficiently high Reynolds numbers, the dissipation scales are much
more separated from the high-$k$ forced scales. In this case a
plateau of $\Pi$ will arise at low wavenumbers: $\Pi(k,t) \approx
\varepsilon_{w,1}$ for $k$ low enough~\cite{mccomb:physics:1990}.
This property is not observed at the computational Reynolds number
considered so far.

In cases specified by Runs~$18 - 19$ we consider the flow at a four
times higher computational Reynolds number. The overall results for
the energy spectra and energy transfer were found to be
qualitatively the same as in the lower Reynolds number cases.
However, a plateau may now be observed in Fig.~\ref{fig:Pi1}, where
we present the transport power for the higher Reynolds number. In
this case the transport power does not decrease below zero in the
forced region. The second forcing band is well separated from the
dissipation region and the transport power in this band is much
larger, approaching a maximum $0.23$ that is near the energy
injection rate~$\varepsilon_{w,2}$.

In this section we have looked at the effect of high-$k$ modulation
of the energy cascading process that leads to an increased energy
dissipation in small scales. This process is supported by an
increased energy transfer to smaller scales via nonlocal triad
interactions. The effect of increased energy rate by the application
of broad-band forcing is seen in the energy transfer and transport
power spectra. In the next section we will look more closely at the
interactions of various scales of motion under the influence of
broad-band forcing by considering the two- and three-mode transfers
$T_{p}$ and $T_{pq}$ introduced in (\ref{eq:Tkpt}) and
(\ref{eq:Tkpqt}).

\section{Two- and three-mode interaction of scales} \label{sec:transfer}

\begin{figure}
\begin{center}
\includegraphics[width=0.45\textwidth]{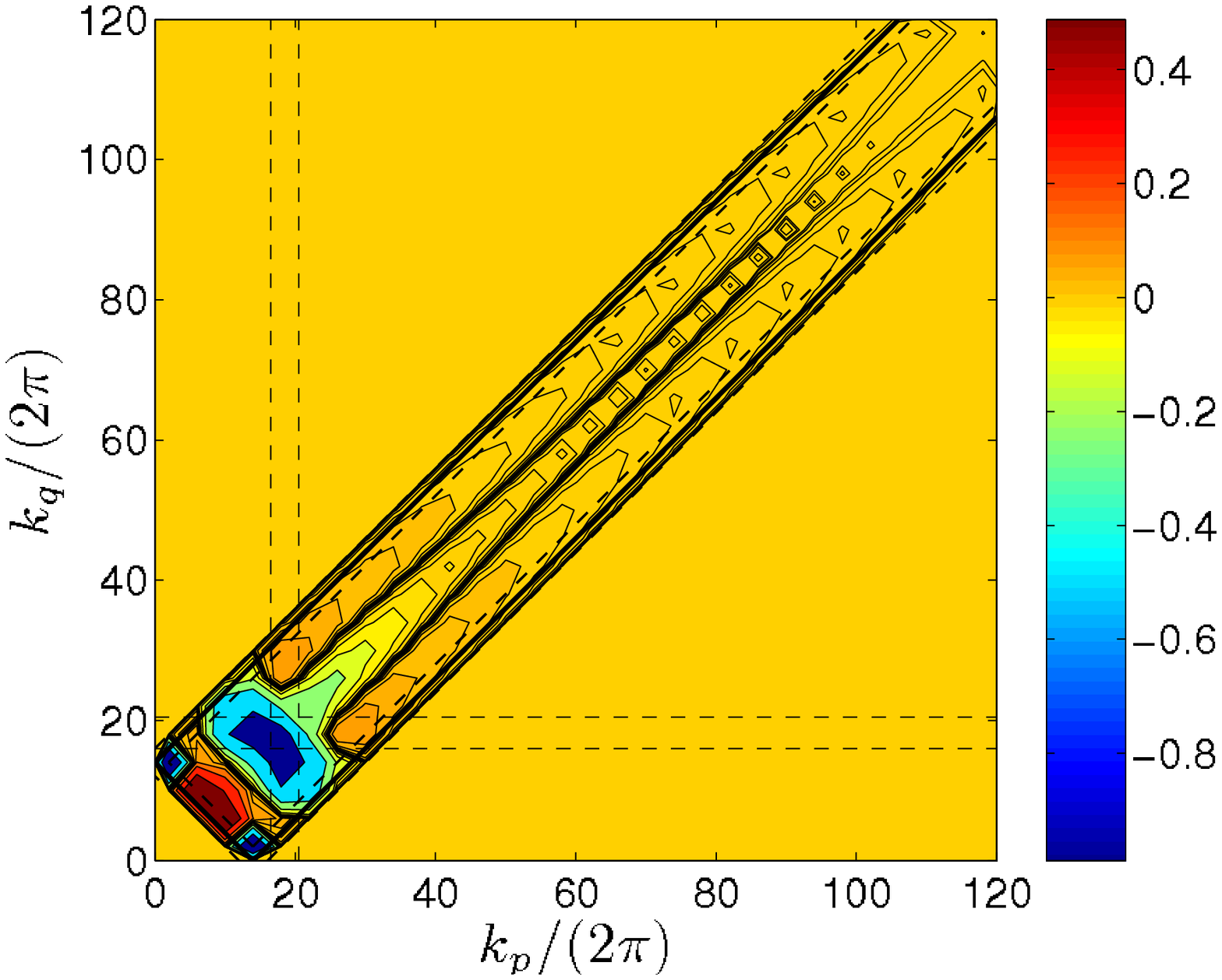}(a)\\
\includegraphics[width=0.45\textwidth]{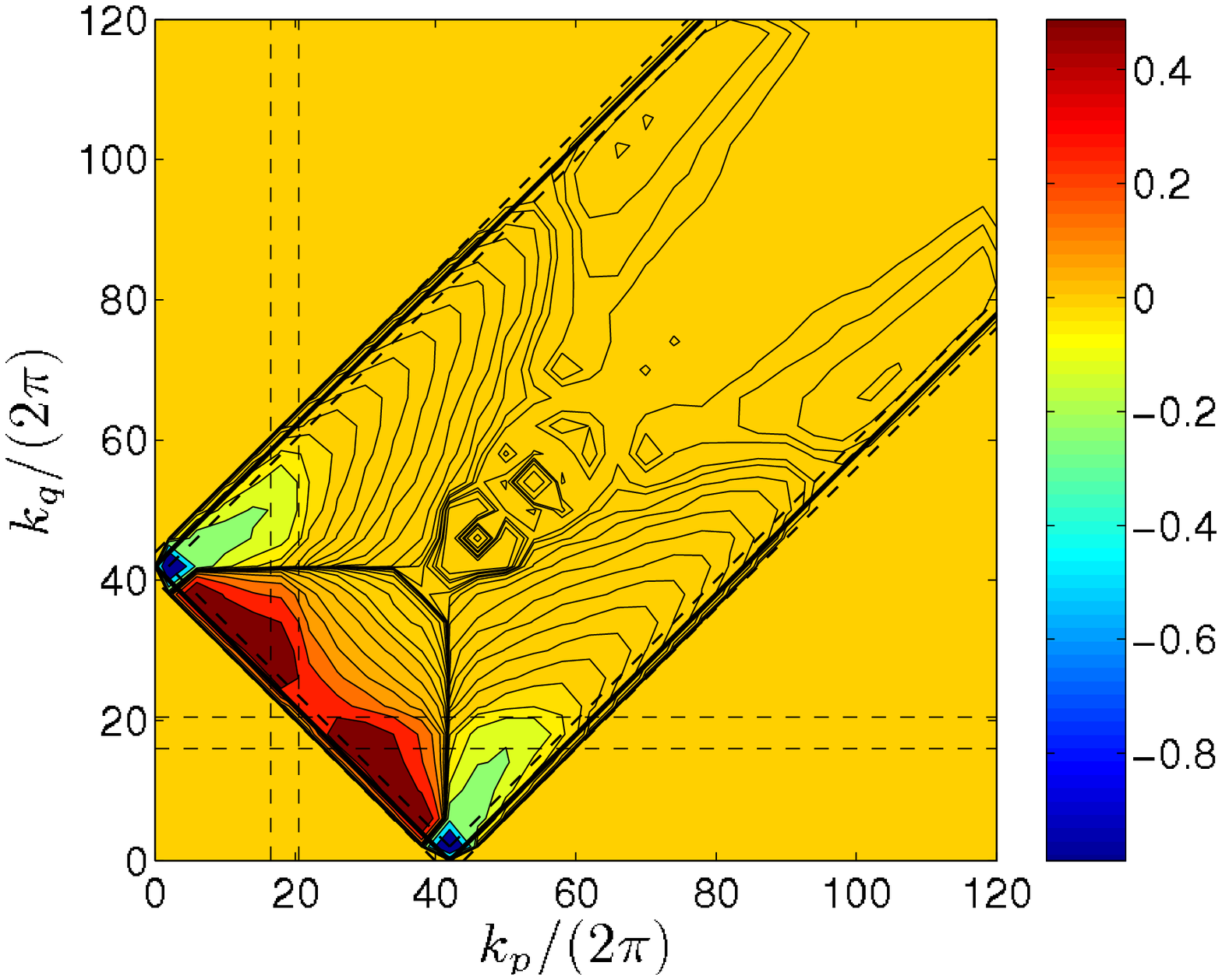}(b)\\
\includegraphics[width=0.45\textwidth]{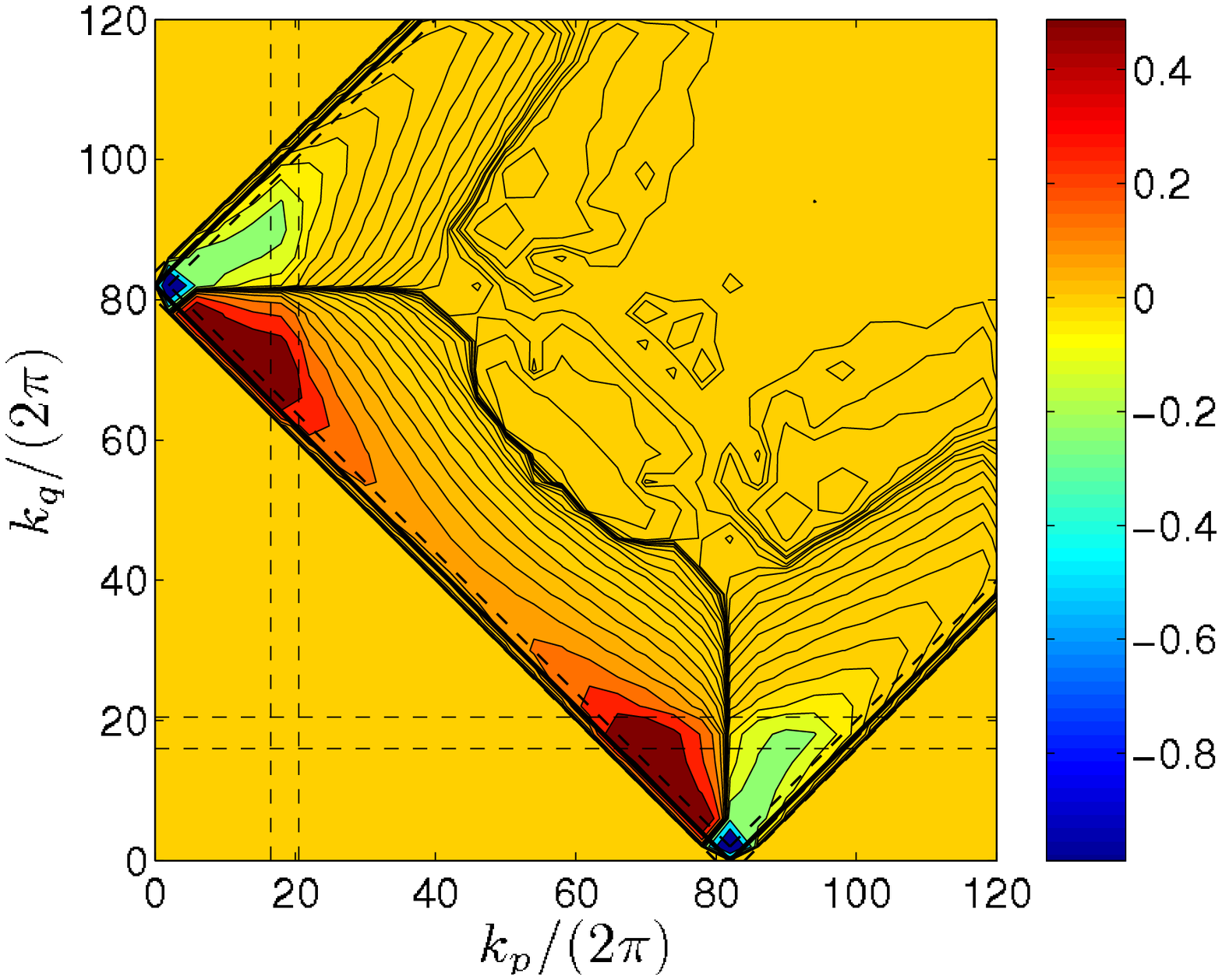}(c)
\end{center}\caption{The normalized triad energy transfer function
$\overline{T}_{pq}(k,k_p,k_q)$ for $k/(2\pi) = 14, 42, 82$ in (a),
(b), (c), respectively. The dashed lines correspond to the lower
($k_1$) and upper ($k_2$) wavenumbers used in the broad-band forcing
at $Re=4243$. The contour levels are $\pm {1/2}^n$, $n=0,\ldots,18$
that are the same for all three pictures.}\label{fig:Tpqk}
\end{figure}

\begin{figure}
\begin{center}
\includegraphics[width=0.45\textwidth]{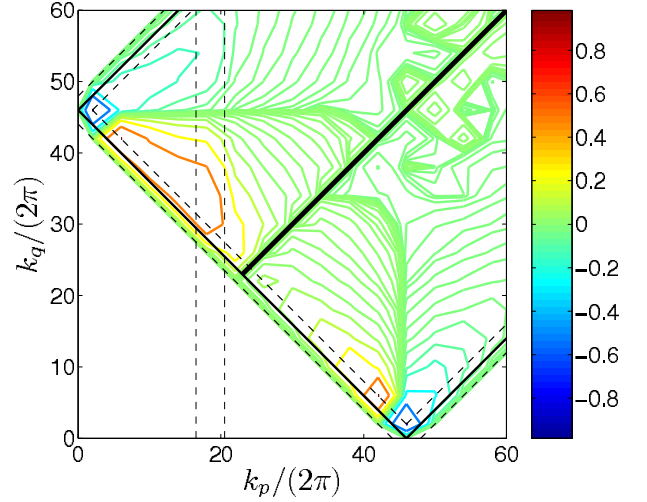}
\caption{The normalized triad energy transfer function
$\overline{T}_{pq}(k,k_p,k_q)$ for $k/(2\pi)=46$ at $Re=4243$. The
upper (bottom) rectangle presents broad-band (large-scale) forced
turbulence respectively. The contour levels are $\pm {1/2}^n$,
$n=0,\ldots,24$.}\label{fig:diff}
\end{center}
\end{figure}

The energy dynamics of turbulent flow is generally discussed in
terms of the transfer of kinetic energy from larger to smaller
scales through nonlinear interactions. The statistical properties of
turbulence are determined by these interactions. In the previous
section we have shown how additional broad-band forcing of inertial
range scales can modify the classical picture of the Kolmogorov
cascade. To investigate the observed turbulence modulation effects
in more detail we consider the underlying two- and three-mode energy
transfer terms in this section. This will clarify to some extent the
changes in the various nonlinear interactions that give rise to the
observed alterations in the spectra and energy transfer.

We start with the three-mode transfer that is averaged in time
$T_{pq}(k,k_p,k_q) = \langle T_{pq}(k,k_p,k_q,t) \rangle_t$ and
split this term into its positive and negative parts:
\begin{equation} \label{eq:Tplusmin}
T_{pq}(k,k_p,k_q)=T^{-}_{pq}(k,k_p,k_q)+T^{+}_{pq}(k,k_p,k_q),
\end{equation}
in which
\begin{equation} \label{eq:Tplus}
T^{+}_{pq}(k,k_p,k_q) = \left\{
\begin{gathered}
  T_{pq}(k,k_p,k_q) ~~~~{\text{if}}~~~~T_{pq}(k,k_p,k_q) \geq 0,  \hfill \\
  0~~~~~~~~~~~~~~~~~~~~{\text{otherwise}}, \hfill \\
\end{gathered}  \right.
\end{equation}
with a similar definition for the negative part:
\begin{equation}\label{eq:Tmin}
T^{-}_{pq}(k,k_p,k_q) = \left\{
\begin{gathered}
  T_{pq}(k,k_p,k_q) ~~~~{\text{if}}~~~~T_{pq}(k,k_p,k_q) < 0,  \hfill \\
  0~~~~~~~~~~~~~~~~~~~~{\text{otherwise}}. \hfill \\
\end{gathered}  \right.
\end{equation}
In terms of these contributions we examine the normalized triad
energy transfer
\begin{equation}
\overline{T}_{pq}(k,k_p ,k_q) = \frac{T^{-}_{pq}(k,k_p,k_q)}{T_{\min}(k)} +
\frac{T^{+}_{pq}(k,k_p,k_q )}{T_{\max}(k)},
\end{equation}
where
\begin{eqnarray}\label{eq:Tminmax}
T_{\min } (k) &=& -\min_{k_{p},k_{q}} \left( {T^ - _{pq} (k,k_p ,k_q
)} \right), \nonumber \\ T_{\max } (k) &=& \max_{k_{p},k_{q}} \left(
{T^ + _{pq} (k,k_p ,k_q )} \right).
\end{eqnarray}
Through the scaling of $T^ - _{pq}$ and $T^+ _{pq}$ with $T_{\min}$
and $T_{\max}$ respectively, the normalized transfer is well suited
to characterize the overall structure of the three-mode transfer
function, even in cases in which the order of magnitude of $T_{pq}$
varies considerably. The normalized energy transfer
$\overline{T}_{pq}(k,k_p ,k_q)$ is plotted in Fig.~\ref{fig:Tpqk}
for three different wavenumbers $k/(2\pi)=14, 42, 82$, based on
Run~$19$ in which $R_\lambda \cong 75$. The three $k$-values that
are selected correspond to wavenumbers below the forced region
($k/(2\pi)=14$) or to wavenumbers that are considerably larger. Such
contour maps for $\overline{T}_{pq}$ can also be found in
\cite{ohkitani:triad:1991} for the case of large-scale forced
turbulence. For completeness, we also presented the results from
such large-scale forced turbulence (Run~$18$) comparing these
directly to the broad-band forced turbulence (Run~$19$). This
contour map is shown in Fig.~\ref{fig:diff} for $k/(2 \pi)=46$.

The strongest interactions are observed for modes with wavenumbers
between the largest forced scales and the high-$k$ forced region as
can be seen in Fig.~\ref{fig:Tpqk}(a). As in the case of large-scale
forcing only we observe very strong interactions between Fourier
modes of considerably different scales. These are located in the
corners of the rectangular domains in the $k_{p}$-$k_{q}$ plane.
Distant interactions are well separated from the origin in these
figures. Their contribution to the transfer is seen to be very
small, as also noticed earlier in the
literature~\cite{ohkitani:triad:1991}. The change of sign in the
transfer function that occurs at $k_p=k$ and $k_q=k$ respectively on
the $k_p$-$k_q$ planes indicates that in this region the energy is
mainly transferred to higher~$k$.

The most efficient transfer takes place between two wavevectors of
similar size and one of quite different size as seen in the corners
of the rectangular area in Fig.~\ref{fig:Tpqk}. This is in agreement
with  previous numerical experiments reported by various
authors~\cite{domaradzki:local:1990,ohkitani:triad:1991,zhou:advances:1998}.
However, compared to the case of large-scale forcing only, we now
observe quite extended, highly energetic interactions with the
high-$k$ forced region. The second forced band causes regions with
high intensity of interactions to be much wider compared to the case
of large-scale forcing only. This is visible directly in
Fig.~\ref{fig:diff}. The regions with positive and negative transfer
are extended from the corners to the wavenumber regions where the
actual application of forcing in the second band occurs. The energy
is exchanged predominantly between scales that are more separated
than in case of the large-scale forced flow where the dominant
interactions occur only in the corners. This is a clear indication
of the stronger non-local interactions, mentioned earlier.

\begin{figure}
\begin{center}
\includegraphics[width=0.45\textwidth]{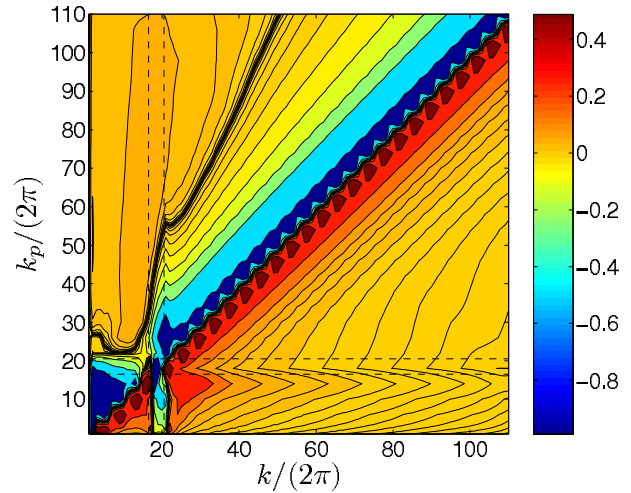}
\caption{The normalized triad energy transfer function
$\overline{T}_{p}(k,k_p)$ at $Re=4243$ (Run~$19$). The contour
levels are $\pm {1/2}^n$, $n=0,\ldots,18$. The supplementary forced
region $k_1 <k \leq k_2$ denoted with dashed
lines.}\label{fig:Tkkp0}
\end{center}
\end{figure}

\begin{figure}
\begin{center}
\includegraphics[width=0.45\textwidth]{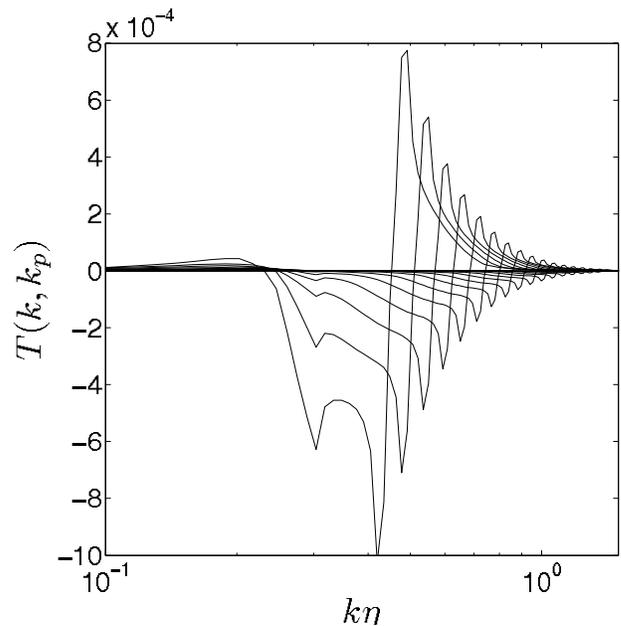}
\end{center}\caption{The triad energy transfer function ${T}_{p}(k,k_p)$ at $k_p/(2\pi) =
30, 34,\ldots, 94$ and $Re=4243$ (Run $19$).}\label{fig:Tkkp2}
\end{figure}

\begin{figure}
\begin{center}
\includegraphics[width=0.45\textwidth]{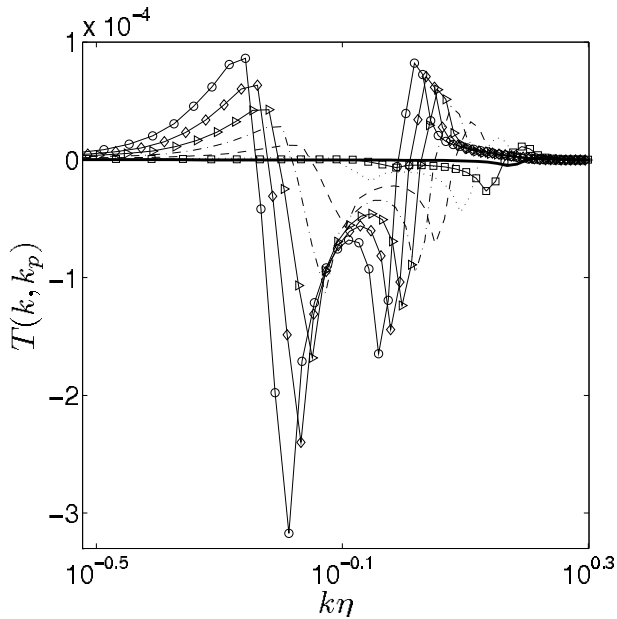}
\end{center}\caption{The triad energy transfer function ${T}_{p}(k,k_p)$ at $k_p/(2\pi) =
30$ and $Re =1061$. Large-scale forcing (solid) with additional
second band forced at $\varepsilon_{w,2}=0.07, 0.15, 0.30, 0.45,
0.60, 0.75, 0.90$  denoted as $\square$, dotted, dashed,
dash-dotted, $\triangleright$, $\diamond$, $\circ$,
respectively.}\label{fig:Tkkp1}
\end{figure}

For further clarification of energy transfer processes we turn to
the time-averaged two-mode energy transfer $T_{p}(k,k_p)=\langle
T_{p}(k,k_p) \rangle_t$, which gives information about the
interactions involving a sum over all $k_q$ wavenumbers at fixed $k$
and $k_{p}$. The sum involves all $k_q$ wavenumbers that are
constrained by the triadic interactions, i.e., their length may vary
between $|\mathbf{k}-\mathbf{p}|$ and $|\mathbf{k}+\mathbf{p}|$. We
normalized the two-mode transfer function $T_{p}(k,k_p)$ in a
similar manner as $T_{p}(k,k_p,k_q)$:
\begin{equation} \label{eq:Tpnorm}
\overline{T}_{p}(k,k_p) = \frac{T^{-}_{p}(k,k_p)}{T_{\min}(k)} +
\frac{T^{+}_{p}(k,k_p )}{T_{\max}(k)},
\end{equation}
where $T^{\pm}_{p}$, $T_{\min}$ and $T_{\max}$ are defined in terms
of $T_{p}$ in a manner analogous to the definitions in
(\ref{eq:Tplusmin}), (\ref{eq:Tplus}), (\ref{eq:Tmin}) and
(\ref{eq:Tminmax}). In Fig.~\ref{fig:Tkkp0} we plotted the contour
map of $\overline{T}_{p}(k,k_p)$.  For larger wavenumbers this
quantity was found to look quite similar to the case of large-scale
forced turbulence. The two-mode transfer function changes sign from
negative to positive at $k=k_p$ indicating a downward energy flow.
Comparing this to the large-scale forced turbulence we observe ({\it
i})~strong influence of forcing in the regions where it is applied
(denoted with dashed lines), ({\it ii})~extended negative energy
transfer region with comparatively high magnitude above the $k=k_p$
line, ({\it iii})~amplification of the backward energy transfer
indicated by the positive region for small $k$ and large $k_p$. This
region is separated from the intense negative energy transfer region
by the indicated accumulation of contour lines above the $k=k_p$
line appearing as the curved black line.

A more quantitative overview is plotted in Fig.~\ref{fig:Tkkp2}
displaying the two-mode transfer function in the range $k_p/(2\pi) =
30, 34, \ldots, 94$. This clearly shows the cascading character of
the energy flow from larger to smaller scales in the system. The
modification due to the high-$k$ forcing expresses itself by the
sequence of one slightly positive, two quite negative and one quite
positive local extrema. The intensity of the energy transfer
decreases with increasing wavenumbers as less energy needs to be
transferred. This corresponds directly to the magnitude of
$T_{\min}(k)$ and $T_{\max}(k)$ used in the normalization of
$T_p(k,k_p)$ (\ref{eq:Tpnorm}). The part in which the transfer is
negative is much wider in the broad-band forced case compared to the
large-scale forced turbulence results.

We conclude by considering the effect of varying the forcing
strength $\varepsilon_{w,2}$ at a characteristic wavenumber
$k_p/(2\pi)=30$ on the two-mode energy transfer function
$T_{p}(k,k_p)$. This is shown in Fig.~\ref{fig:Tkkp1}. In the
large-scale forced case at $\varepsilon_{w,2}=0$ the transfer is
very small compared to the cases in which the high-$k$ forcing is
active. In addition, the effect is very localized (solid line in
Fig.~\ref{fig:Tkkp1}). The forcing in the high-$k$ band completely
changes this behavior. The intensity of the energy transfer is
directly related to the value of $\varepsilon_{w,2}$. Additional
extrema appear in the two-mode transfer function.  The high-$k$
forced cases display two pairs in which a negative minimum is
combined with a positive maximum, while large scale forcing only
yields one such combination. Correspondingly, the min-max pair at
high $k$ is associated with the large scale forcing in
$\mathbb{S}_{1}$ while the min-max pair at lower $k$ originates from
the additional forcing in the second band. We also investigated
three-band forcing and observed further peaks in the energy transfer
spectra.

\section{Concluding remarks} \label{sec:conc}

We performed direct numerical simulations of broad-band forced
turbulence to explore accumulated effects on the time-averaged
energy transfer in isotropic homogeneous turbulence. Using
broad-band forcing based on a recently proposed mathematical model
for a fractal stirrer~\cite{mazzi:fractal:2004} we have shown how
the application of such forcing modulates turbulence both
qualitatively and quantitatively. The modulation is similar
to that observed in experiments based on flows through porous media
or canopies. Specifically the perturbation of a flow arising from
the contact with complex physical boundaries enhances the
dissipation and causes an abrupt energy drain from large to small
scales. This aspect of simultaneous perturbation of a flow on a
spectrum of length-scales is retained in the cases studied here.

We found that broad-band forcing that perturbs a turbulent flow at
smaller scales enhances non-local triad interactions and alters the
detailed cancellation processes that occur in the traditional
large-scale forced flows. This leads to non-local modifications in
the energy transfer spectrum and the energy distribution among
scales. We verified this by partitioning the nonlinear term in the
Navier--Stokes equations in terms of different triad contributions
to the total transfer function. The energy transport power is found
to be enhanced in the spectral region in between the large-scale and
the high-$k$ forced bands. This characteristic may be influenced via
the control parameters of the applied forcing, i.e., its strength
and extent of agitated scales, and allows optimizing transport
processes of turbulent flows.

Future study will involve the examination of the consequences of
forcing in the physical space context. We will investigate the
geometrical statistics of broad-band forced turbulence looking at
the interactions of strain and vorticity and their modulation by the
applied forcing. This may help understanding which physical
processes are responsible for the observed modulations and how to
exploit this to enhance physical space mixing.

\appendix
\section*{Appendix}

The main parameters of the simulations are collected in
Table~\ref{tab:pa}.  The corresponding statistics of the velocity
fields are summarized in Table~\ref{tab:da}. The quantities compiled
in Table \ref{tab:da} are the Kolmogorov dissipation wavenumber
$k_d$ which is the inverse of the Kolmogorov length-scale $\eta$,
the product $k_{\max} \eta$, the Taylor microscale
$\lambda=(5\widehat{E}/\sum_{\mathbf{k}}
{k^{2}{E}(\mathbf{k},t)})^{1/2}$, the Taylor-microscale Reynolds
number $R_{\lambda}=\lambda u'/\nu$, the integral length-scale $L=
3\pi /(4\widehat{E}) \sum_{\mathbf{k}} {k^{-1}{E}(\mathbf{k},t)}$,
the integral Reynolds number $R_{L}= L u'/\nu$, the r.m.s velocity
$u'=({2\widehat{E}/3})^{1/2}$, the energy dissipation rate
$\varepsilon=\sum_{\mathbf{k}} {2\nu k^2 {E}(\mathbf{k},t)}$, the
eddy-turnover time $\tau =L /u'$ and the skewness $S = 2/35 \left(
\lambda / u' \right)^3 \sum_{\mathbf{k}} {k^2 {T}(\mathbf{k},t)}$.
All these quantities in Table \ref{tab:da} are time-averaged
$\langle \cdot \rangle_{t}$ as described in Sec.~\ref{simu}.

We also checked that the alteration of the cascading process
caused by the high-$k$ forcing does not influence the isotropy of
the flow field. A measure of isotropy was suggested in
\cite{curry:order:1984} given by: $I^2(t)=\psi_1(t)/\psi_2(t)$ where
$\psi_1(t)=\left\langle |\mathbf e_1(\mathbf{k}) \mathbf
{u}(\mathbf{k},t)|^2 \right\rangle$, $\psi_2(k,t)=\left\langle
|\mathbf e_2(\mathbf{k}) \mathbf {u}(\mathbf{k},t)|^2 \right\rangle
$ are the kinetic energy along the components of two orthogonal
solenoidal unit vectors $\mathbf e_1(\mathbf{k}) = \mathbf{k} \times
\mathbf z(\mathbf{k}) / |\mathbf{k} \times \mathbf z (\mathbf{k})|$,
$\mathbf e_2(\mathbf{k}) = \mathbf{k} \times \mathbf
e_1(\mathbf{k})/ |\mathbf{k} \times \mathbf e_1 (\mathbf{k})|$ where
$\mathbf{z}(\mathbf{k})$ is a randomly oriented unit vector.  The
operator $\langle\cdot\rangle$ denotes averaging over these random
unit vectors. For isotropic turbulence one can expect to find $I=1$,
i.e., $\psi_1=\psi_2$ which was confirmed to close approximation in
all simulations. Deviations from the expected value for $I$ were
found to be of the order of $1\%$. \bigskip

\begin{table}
\begin{center}
\begin{tabular}{||c|c|c|c||c|c|c|c|}
  \hline
  Run &  $\varepsilon_{w}$ & $m$ & $p$ & Run &  $\varepsilon_{w}$ & $m$ & $p$\\
  \hline
  $1$     &  $0.15^{\star}$& $-$ & $-$  &   $$  & $$     & $$ & $$\\
  \hline
  $2$     &  $0.07$    & $17$ & $20$  &  ${8}$  & $0.15$     & $17$ & $24$\\
  $3$     &  $0.30$    & $17$ & $20$  &  ${9}$  & $0.30$     & $17$ & $24$\\
  $4$     &  $0.45$    & $17$ & $20$  &  ${10}$  & $0.45$     & $17$ & $24$\\
  $5$     &  $0.60$    & $17$ & $20$  &  ${11}$  & $0.60$     & $17$ & $24$\\
  $6$     &  $0.75$    & $17$ & $20$  &  ${12}$  & $0.75$     & $17$ & $24$\\
  $7$     &  $0.90$    & $17$ & $20$  &  ${13}$  & $0.90$     & $17$ & $24$\\
  \hline
  ${14}$     &  $0.15$    & $5$ & $8$   &   ${18}$ & $0.15^{\star}$ & $-$ & $-$\\
  ${15}$     &  $0.15$    & $9$ & $12$  &   ${19}$ & $0.30$         & $17$ & $20$\\
  ${16}$     &  $0.15$    & $17$& $20$  &   $$ & $$            & $$ & $$\\
  ${17}$     &  $0.15$    & $25$& $28$  &   $$ & $$            & $$ & $$\\
  \hline
\end{tabular}
\end{center}
\caption{Direct numerical simulation parameters using a~resolution
of $N=128$ and $Re=1061$ in Runs $1 - 17$, and a~resolution of
$N=256$ at $Re=4243$ in Runs ${18} - {19}$. The cases with
large-scale forcing only are denoted by ${\star}$. In this table
$\varepsilon_{w}$ denotes the energy input-rate in the high-$k$
band, except Run~$1$ and ${18}$ in which it corresponds to the
energy input-rate in $\mathbb{S}_{1}$. Moreover, $m$ and $p$
characterize the spectral support of the high-$k$ band
$\mathbb{K}_{m,p}$.} \label{tab:pa}
\end{table}

\begin{table}
\begin{center}
\begin{tabular}{|c||c|c|c|c|c|c|c|c|c|c|c|c}
  \hline
  Run        & $k_d$ & $k_{\max}\eta$ & $\lambda$ & $R_{\lambda}$ & $L$ & $R_{L}$ &  $u'$ & $\varepsilon$ & $\tau$ & $S$ \\
  \hline
  $1$      & $116$ & $3.26$ & $0.123$& $52$ & $0.23$ & $97$  & $0.40$ & $0.15$ & $0.57$ & $0.49$\\
  \hline 
  $2$      & $130$ & $2.91$ & $0.100$& $43$ & $0.23$ & $100$ & $0.41$ & $0.24$ & $0.56$ & $0.35$\\
  $3$      & $156$ & $2.42$ & $0.069$& $30$ & $0.23$ & $98$  & $0.41$ & $0.50$ & $0.54$ & $0.21$\\
  $4$      & $168$ & $2.25$ & $0.061$& $27$ & $0.23$ & $101$ & $0.42$ & $0.66$ & $0.54$ & $0.17$\\
  $5$      & $178$ & $2.12$ & $0.056$& $25$ & $0.22$ & $102$ & $0.43$ & $0.83$ & $0.52$ & $0.15$\\
  $6$      & $186$ & $2.03$ & $0.051$& $23$ & $0.22$ & $102$ & $0.43$ & $1.00$ & $0.52$ & $0.14$\\
  $7$      & $193$ & $1.95$ & $0.048$& $22$ & $0.22$ & $100$ & $0.43$ & $1.16$ & $0.50$ & $0.13$\\
  \hline 
  ${8}$    & $140$ & $2.68$ & $0.085$& $37$ & $0.23$ & $99$  & $0.41$ & $0.33$ & $0.56$ & $0.29$ \\
  ${9}$    & $156$ & $2.41$ & $0.070$& $31$ & $0.23$ & $102$ & $0.41$ & $0.50$ & $0.56$ & $0.22$ \\
  ${10}$   & $169$ & $2.24$ & $0.060$& $26$ & $0.22$ & $98$  & $0.41$ & $0.68$ & $0.54$ & $0.18$ \\
  ${11}$   & $178$ & $2.11$ & $0.054$& $24$ & $0.22$ & $100$ & $0.42$ & $0.85$ & $0.53$ & $0.16$ \\
  ${12}$   & $187$ & $2.02$ & $0.049$& $22$ & $0.22$ & $97$  & $0.42$ & $1.03$ & $0.52$ & $0.15$ \\
  ${13}$   & $194$ & $1.94$ & $0.046$& $21$ & $0.21$ & $96$  & $0.42$ & $1.20$ & $0.50$ & $0.13$ \\
  \hline 
  ${14}$   & $138$ & $2.73$ & $0.090$& $39$ & $0.22$ & $96$  & $0.41$ & $0.30$ & $0.52$ & $0.44$\\
  ${15}$   & $138$ & $2.72$ & $0.089$& $39$ & $0.23$ & $102$ & $0.42$ & $0.31$ & $0.56$ & $0.39$\\
  ${16}$   & $140$ & $2.69$ & $0.084$& $36$ & $0.22$ & $96$  & $0.40$ & $0.33$ & $0.55$ & $0.28$\\
  ${17}$   & $143$ & $2.64$ & $0.082$& $36$ & $0.23$ & $98$  & $0.41$ & $0.35$ & $0.56$ & $0.22$\\
  \hline 
  ${18}$   & $325$ & $2.32$ & $0.065$& $115$ & $0.21$ & $368$  & $0.42$ & $0.15$ & $0.50$ & $0.51$ \\
  ${19}$   & $432$ & $1.74$ & $0.040$& $75$ & $0.21$ & $394$  & $0.44$ & $0.46$ & $0.47$ & $0.38$ \\
  \hline
\end{tabular}
\end{center}
\caption{Direct numerical simulations statistics of the different
cases studied.} \label{tab:da}
\end{table}

\section*{Acknowledgments}
This work is part of the research program ``Turbulence and its role
in energy conversion processes'' of the Foundation for Fundamental
Research of Matter (FOM). The authors wish to thank NCF (Dutch
foundation for National Computing Facilities) for supporting the
computations. These were executed at SARA Computing and Networking
Services in Amsterdam.

\bibliographystyle{unsrt}
\bibliography{transfer_pre}

\begin{thebibliography}{10}

\bibitem{mccomb:physics:1990}
W.~D. McComb.
\newblock {\em The Physics of Fluid Turbulence}.
\newblock Oxford University Press, 1990.

\bibitem{zhou:advances:1998}
Y.~Zhou and Ch.~G. Speziale.
\newblock Advances in the fundamental aspects of turbulence: Energy transfer,
  interacting scales, and self-preservation in isotropic decay.
\newblock {\em Appl. Mech. Rev.}, 51(4):267--301, 1998.

\bibitem{siggia:numerical:1981}
E.~D. Siggia.
\newblock Numerical study of small-scale intermittency in three-dimensional
  turbulence.
\newblock {\em J. Fluid Mech.}, 107:375--406, 1981.

\bibitem{kerr:higher:1985}
R.~M. Kerr.
\newblock Higher-order derivative correlations and the alignment of small-scale
  structures in isotropic numerical turbulence.
\newblock {\em J. Fluid Mech.}, 153:31--58, 1985.

\bibitem{eswaran:examination:1988}
V.~Eswaran and S.~B. Pope.
\newblock An examination of forcing in direct numerical simulations of
  turbulence.
\newblock {\em Comput. Fluids}, 16:257--278, 1988.

\bibitem{chen:high:1992}
S.~Chen and X.~Shan.
\newblock High-resolution turbulent simulations using the {Connection
  Machine-2}.
\newblock {\em Comput. Phys.}, 6(6):643--646, 1992.

\bibitem{jimenez:structure:1993}
J.~Jimenez, A.~A. Wray, P.~G. Saffman, and R.~S. Rogallo.
\newblock The structure of intense vorticity in isotropic turbulence.
\newblock {\em J. Fluid Mech.}, 255:65--90, 1993.

\bibitem{ghosal:dynamic:1995}
S.~Ghosal, T.~S. Lund, P.~Moin, and K.~Akselvoll.
\newblock A dynamic localization model for large-eddy simulation of turbulent
  flows.
\newblock {\em J. Fluid Mech.}, 286:229--255, 1995.

\bibitem{machiels:predictability:1997}
L.~Machiels.
\newblock Predictability of small-scale motion in isotropic fluid turbulence.
\newblock {\em Phys. Rev. Lett.}, 79(18):3411--3414, 1997.

\bibitem{overholt:deterministic:1998}
M.~R. Overholt and S.~B. Pope.
\newblock A deterministic forcing scheme for direct numerical simulations of
  turbulence.
\newblock {\em Comput. Fluids}, 27(1):11--28, 1998.

\bibitem{kaneda:energy:2003}
Y.~Kaneda, T.~Ishihara, M.~Yokokawa, K.~Itakura, and A.~Uno.
\newblock Energy dissipation rate and energy spectrum in high resolution direct
  numerical simulations of turbulence in a periodic box.
\newblock {\em Phys. Fluids}, 15(2):L21--L24, 2003.

\bibitem{kolmogorov:local:1941}
A.~N. Kolmogorov.
\newblock The local structure of turbulence in incompressible viscous fluids at
  very large {Reynolds} numbers.
\newblock {\em C.R. Acad. Sci. URSS}, 30:301--305, 1941.

\bibitem{zhou:degreees:1993}
Y.~Zhou.
\newblock Degrees of locality of energy transfer in the inertial range.
\newblock {\em Phys. Fluids A}, 5(5):1092--1094, 1993.

\bibitem{domaradzki:analysis:1987}
J.~A. Domaradzki, R.~W. Metcalfe, R.~S. Rogallo, and J.~J. Riley.
\newblock Analysis of subgrid-scale eddy viscosity with use of results from
  {Direct Numerical Simulations}.
\newblock {\em Phys. Rev. Lett.}, 58(6):547--550, 1987.

\bibitem{domaradzki:analysis:1988}
J.~A. Domaradzki.
\newblock Analysis of energy transfer in direct numerical simulations of
  isotropic turbulence.
\newblock {\em Phys. Fluids}, 31(10):2747--2749, 1988.

\bibitem{domaradzki:local:1990}
J.~A. Domaradzki and R.~S. Rogallo.
\newblock Local energy transfer and nonlocal interactions in homogeneous,
  isotropic turbulence.
\newblock {\em Phys. Fluids A}, 2(3):413--426, 1990.

\bibitem{domaradzki:nonlocal:1992}
J.~A. Domaradzki.
\newblock Nonlocal triad interactions and the dissipation range of isotropic
  turbulence.
\newblock {\em Phys. Fluids A}, 4(9):2037--2045, 1992.

\bibitem{domaradzki:energy:1994}
J.~A. Domaradzki, W.~Lui, C.~Hartel, and L.~Kleiser.
\newblock Energy transfer in numerically simulated wall-bounded turbulent
  flows.
\newblock {\em Phys. Fluids}, 6(4):1583--1599, 1994.

\bibitem{ohkitani:triad:1991}
K.~Ohkitani and S.~Kida.
\newblock Triad interactions in a forced turbulence.
\newblock {\em Phys. Fluids A}, 4(4):794--802, 1992.

\bibitem{waleffe:nature:1991}
F.~Waleffe.
\newblock The nature of triad interactions in homogenous turbulence.
\newblock {\em Phys. Fluids A}, 4(2):350--363, 1992.

\bibitem{kuczaj:mixing:2006}
A.~K. Kuczaj and B.~J. Geurts.
\newblock Mixing in manipulated turbulence.
\newblock {\em J. Turbul.}, submitted.

\bibitem{mccomb:drag:1981}
W.~D. McComb and K.~T.~J. Chan.
\newblock Drag reduction in fibre suspension.
\newblock {\em Nature}, 292:520--522, 1981.

\bibitem{mccomb:laser:1985}
W.~D. McComb and K.~T.~J. Chan.
\newblock {Laser-Doppler} anemometer measurements of the turbulent structure in
  drag-reducing fibre suspensions.
\newblock {\em J. Fluid Mech.}, 152:455--478, 1985.

\bibitem{boomsma:metal:2002}
K.~Boomsma, D.~Poulikakos, and F.~Zwick.
\newblock Metal foams as compact high performance heat exchangers.
\newblock {\em Mech. Mater.}, 35:1161--1176, 2003.

\bibitem{finnigan:turbulence:2000}
J.~Finnigan.
\newblock Turbulence in plant canopies.
\newblock {\em Ann. Rev. Fluid Mech.}, 32:519--571, 2000.

\bibitem{pouquet:turbulence:1983}
A.~Pouquet, U.~Frisch, and J.~P. Chollet.
\newblock Turbulence with a spectral gap.
\newblock {\em Phys. Fluids}, 26(4):877--880, 1983.

\bibitem{mazzi:fractal:2004}
B.~Mazzi and J.~C. Vassilicos.
\newblock Fractal generated turbulence.
\newblock {\em J. Fluid Mech.}, 502:65--87, 2004.

\bibitem{yeung:response:1991}
P.~K. Yeung and J.~G. Brasseur.
\newblock The response of isotropic turbulence to isotropic and anisotropic
  forcing at the large scales.
\newblock {\em Phys. Fluids A}, 3(5):884--897, 1991.

\bibitem{brasseur:interscale:1994}
J.~G. Brasseur and Ch-H. Wei.
\newblock Interscale dynamics and local isotropy in high {Reynolds} number
  turbulence within triadic interactions.
\newblock {\em Phys. Fluids}, 6(6):842--870, 1994.

\bibitem{zhou:interacting:1993}
Y.~Zhou.
\newblock Interacting scales and energy transfer in isotropic turbulence.
\newblock {\em Phys. Fluids A}, 5(10):2511--2524, 1993.

\bibitem{zhou:scale:1996}
Y.~Zhou, P.~K. Yeung, and J.~G. Brasseur.
\newblock Scale disparity and spectral transfer in anisotropic numerical
  turbulence.
\newblock {\em Phys. Rev. E}, 53(1):1261--1264, 1996.

\bibitem{suzuki:modification:1999}
Y.~Suzuki and Y.~Nagano.
\newblock Modification of turbulent helical/nonhelical flows with small-scale
  energy input.
\newblock {\em Phys. Fluids}, 11(11):3499--3511, 1999.

\bibitem{moin:feedback:1994}
P.~Moin and T.~Bewley.
\newblock Feedback control of turbulence.
\newblock {\em Appl. Mech. Rev.}, 47:S3, 1994.

\bibitem{geurts:elements:2004}
B.~J. Geurts.
\newblock {\em Elements of Direct and Large-Edddy Simulation}.
\newblock R.T. Edwards, 2004.

\bibitem{canuto:spectral:1988}
C.~Canuto, M.~Hussaini, A.~Quarteroni, and T.~Zang.
\newblock {\em Spectral Methods in Fluid Dynamics}.
\newblock Springer Verlag (Berlin and New York), 1988.

\bibitem{yeung:dynamic:1995}
P.~K. Yeung, J.~G. Brasseur, and Q.~Wang.
\newblock Dynamics of direct large-small scale couplings in coherently forced
  turbulence: concurrent physical- and fourier-space views.
\newblock {\em J. Fluid Mech.}, 283:43--95, 1995.

\bibitem{mccomb:conditional:2001}
W.~D. McComb, A.~Hunter, and C.~Johnston.
\newblock Conditional mode-elimination and the subgrid-modeling problem for
  isotropic turbulence.
\newblock {\em Phys. Fluids}, 13(7):2030--2044, 2001.

\bibitem{curry:order:1984}
J.~H. Curry, J.~R. Herring, J.~Loncaric, and S.~A. Orszag.
\newblock Order and disorder in two- and three dimensional {B{\'e}nard}
  convection.
\newblock {\em J. Fluid Mech.}, 147:1--38, 1984.

\end{thebibliography}

\end{document}